\def\year{2022}\relax
\documentclass[letterpaper]{article} 
\usepackage{aaai22}  
\usepackage{times}  
\usepackage{helvet}  
\usepackage{courier}  
\usepackage[hyphens]{url}  
\usepackage{graphicx} 
\usepackage{natbib}  
\usepackage{caption} 
\DeclareCaptionStyle{ruled}{labelfont=normalfont,labelsep=colon,strut=off} 
\usepackage[utf8]{inputenc} 
\usepackage[T1]{fontenc}    
\usepackage{hyperref}       
\usepackage{url}            
\usepackage{booktabs}       
\usepackage{amsfonts}       
\usepackage{nicefrac}       
\usepackage{microtype}      
\usepackage{lipsum}
\usepackage{xcolor}
\usepackage{graphicx}
\usepackage{amsmath}
\usepackage{amsthm}
\usepackage{wrapfig}
\usepackage{array}

\usepackage{amsmath,amsfonts,amssymb}
\usepackage{algorithmic}
\usepackage{textcomp}
\usepackage{booktabs}
\usepackage{enumitem}
\usepackage[ruled, boxed,vlined,linesnumbered]{algorithm2e}
\usepackage{verbatim}
\usepackage{subcaption}
\usepackage{balance}
\usepackage{ifthen}
\usepackage{float}
\usepackage{fancyhdr}
\usepackage{lastpage}
\usepackage{multirow}
\usepackage{wrapfig}

\hypersetup{
  colorlinks   = true, 
  urlcolor     = black, 
  linkcolor    = red, 
  citecolor   = blue 
}

\theoremstyle{definition}
\newtheorem{definition}{Definition}

\newcommand{\scheme}{FDN}

\DeclareMathOperator*{\argmin}{arg\,min}

\title{A Fusion-Denoising Attack on InstaHide with Data Augmentation\footnote{This paper has been accepted by AAAI 2022.}}

\author{Xinjian Luo, Xiaokui Xiao, Yuncheng Wu, Juncheng Liu, Beng Chin Ooi
}
\affiliations{
National University of Singapore\\
\texttt{\{xinjluo, xiaoxk, wuyc, juncheng, ooibc\}@comp.nus.edu.sg}
}

\begin{document}




\maketitle

\begin{abstract}

InstaHide is a state-of-the-art mechanism for protecting private training images, by mixing multiple private images and modifying them such that their visual features are indistinguishable to the naked eye. In recent work, however, Carlini et al.\ show that it is possible to reconstruct private images from the encrypted dataset generated by InstaHide. Nevertheless, we demonstrate that Carlini et al.'s attack can be easily defeated by incorporating data augmentation into InstaHide. This leads to a natural question: is InstaHide with data augmentation secure?
In this paper, we provide a negative answer to this question, by devising an attack for recovering private images from the outputs of InstaHide even when data augmentation is present.
The basic idea is to use a comparative network to identify encrypted images that are likely to correspond to the same private image, and then employ a fusion-denoising network for restoring the private image from the encrypted ones, taking into account the effects of data augmentation.
Extensive experiments demonstrate the effectiveness of the proposed attack in comparison to Carlini et al.'s attack.

\end{abstract}

\section{Introduction}
\begin{figure}[t]
\centering
\includegraphics[width=0.95\columnwidth]{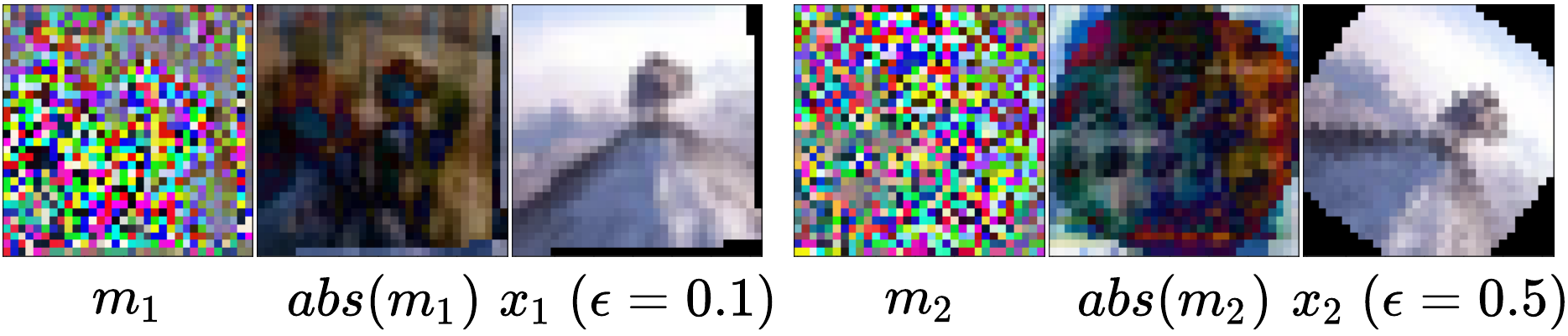}
\caption{Examples generated by InstaHide with data augmentation. $m_i$: the original encryption; $abs(m_i)$: the image after removing all pixel signs of $m_i$; $x_i$: the original private image. $\epsilon$ is defined in the Preliminary section.}
\label{fig-mixegg}
\end{figure}

Collaborative learning~\cite{ yang2019federated, li2020federated, WuCXCO20} is an increasingly popular learning paradigm as it enables multiple data providers to jointly train models without disclosing their private data.
However, recent studies on model inversion attacks~\cite{fredrikson2015model, hitaj2017deep, zhu2020deep, luo2021feature} demonstrate that the training data can be precisely recovered based on the gradients or model parameters shared during collaborative learning. This leads to concerns on the security of existing collaborative learning methods~\cite{huang2020instahide, kairouz2019advances}. 
To mitigate the above concerns, \cite{huang2020instahide} propose a practical scheme, InstaHide, which generates the training datasets by mixing multiple private images into one image~\cite{zhang2017mixup}. The training images produced by InstaHide are called \emph{encryptions} in \cite{huang2020instahide}. Intuitively, InstaHide aims to corrupt the visual features of the private images (as shown in Fig.~\ref{fig-mixegg}) such that the encrypted training images fed into the models are hardly distinguishable by the naked eye, thus eliminating the threats caused by model inversion attacks~\cite{hitaj2017deep, zhu2020deep}. 

Recently, however, \cite{carlini2020attack} propose an attack that can approximately recover the private images encrypted by InstaHide. The main idea of~\cite{carlini2020attack} is to first cluster the encrypted images based on a similarity metric, and then restore one private image
from one cluster of encryptions by factoring out the useless components.
%
Although this attack works well against the InstaHide Challenge dataset~\cite{instachallenge}, there are three main limitations.
\emph{First}, \cite{carlini2020attack} is specially designed for the InstaHide Challenge, where each private image is directly mixed into $T=50$ encryptions. But in applications that $T$ is set to a small number (e.g., 10), the performance of \cite{carlini2020attack} is greatly degraded (as pointed out by one author of InstaHide~\cite{websanj}).
\emph{Second}, the private images could be pre-processed by data augmentation before mixing with other images (this case is included in the source code of InstaHide~\cite{instahidecode} instead of the challenge dataset~\cite{instachallenge}), and  \cite{carlini2020attack} can barely restore distinguishable images.
\emph{Third}, \cite{carlini2020attack} can not precisely restore the original color profiles of the private images,
which would degrade the visual features of the restored images and lead to indistinguishable results.
In this paper, we investigate a more restricted but more practical problem: \emph{how to precisely restore the visual structures and color profiles of a private image from a small number of encryptions generated by InstaHide with data augmentation}?

To address this problem, the general idea is first to determine a set of encryptions that contain the same private image (called \textit{homogeneous encryptions}), then restore the private image based on these homogeneous encryptions. In particular, we view the component produced by irrelevant mixed images in an encryption as noise.
Although the noise pattern is hard to be mathematically formulated because of the randomly nonlinear variations on the mixed pixels introduced by InstaHide, it can be learned effectively by a deep neural network. In this way, we can use a trained network to remove the noise component and accurately restore the color profiles and structures of the private image from a small number ($\ll 50$) of encryptions.


Implementing such a neural network is not trivial. Without careful design, the restored images could be meaningless, as shown in~\cite{huang2020instahide}. The main difficulty is that the private image could be  randomly transformed by geometrical data augmentation before being mixed into multiple encryptions. Since the salient structures of the private image are severely corrupted after being processed by InstaHide (Fig.~\ref{fig-mixegg}), the widely used image registration methods~\cite{zitova2003imgreg, ma2019infrared} that rely on visual features to geometrically align the structures of multiple images are hardly useful. 
Therefore, we have to design an image registration method from scratch to align the salient structures. In addition, we need to handle the case that one image is mixed $\ll 50$ times, in which the pixel-wise optimization method used in \cite{carlini2020attack} can not work because the information provided by the corresponding encoded pixels ($\ll 50$) that are derived from the same private pixel $p$ is not sufficient to recover $p$ (as shown in Fig.~\ref{fig-different-M}). We need to consider a patch-wise restoration method in which the neighboring information of $p$ is used for its restoration.

To overcome these difficulties, we take the attack on InstaHide with data augmentation as an image fusion-denoising task whose inputs are not pre-registered and severely corrupted, and design a registration-fusion-denoising pipeline to handle this task.
We first devise a network component called \textit{image relaxing} to automatically align the severely corrupted private images. Image relaxing can also reduce the noises caused by the structures of other irrelevant mixed images. We further give an extensive analysis of the noise pattern introduced by InstaHide, which inspires us that the corruption levels of the private image can be reflected by the pixel variance. Accordingly, we propose a re-weighting method based on the image variance to pre-process the encryptions before feeding them into the neural network. 
Following these insights, we then design a novel \textbf{F}usion-\textbf{D}enoising \textbf{N}etwork (\scheme) to fuse several homogeneous encryptions into a single encryption and denoise this encryption to recover the private image. 
%
To our knowledge, this is the first work that utilizes a registration-fusion-denoising pipeline to solve the image reconstruction tasks based on the inputs with  not pre-registered and severely corrupted visual features.
We conduct extensive experiments to evaluate the generalization and attack performance of \scheme . The results demonstrate the superior performance of the proposed scheme to~\cite{carlini2020attack}.

\section{Related Work}
\emph{Image fusion} is used to integrate the complementary features of multiple images into one informative image~\cite{zhang2020ifcnn}. Before a  fusion, the images capturing the same scene but in different perspectives should be geometrically aligned, which is known as \textit{image registration}. The traditional registration studies~\cite{ma2019infrared}, which mainly focus on extracting and aligning salient structures, such as edges and corners, are barely useful if the image structures are severely corrupted.
Most fusion studies~\cite{ma2019infrared, zhang2017boundary, liu2015transform, zhang2020ifcnn} assume that the images input to the fusion algorithms are pre-registered, and few of them consider the impact of image noise.
Although a few studies consider joint image fusion and denoising~\cite{li2018joint, liu2020infrared, mei2019simultaneous}, they assume that the visual features of the input images are pre-aligned and not corrupted by the noise, which is not applicable for attacking the InstaHide with data augmentation.

Mixup is proposed as a regularization method for neural network training~\cite{zhang2017mixup}.
Since Mixup can obfuscate the visual features of images, some recent studies
~\cite{fu2019mixup, raynal2020image, zhang2021exploiting} employ it to pre-process the raw training data for privacy-preserving.
However, \cite{huang2020instahide} demonstrate that one private image could be simply restored by averaging those mixup images containing it. Accordingly, \cite{huang2020instahide} propose InstaHide to enhance the security of Mixup.
But~\cite{carlini2020attack} devise an attack that can restore distinguishable images from the InstaHide Challenge dataset~\cite{instachallenge} by minimizing the norm of the noise component.
Nevertheless, \cite{carlini2020attack} is specifically designed for the challenge dataset, which is not general and can be easily defeated by incorporating data augmentation into InstaHide.
On the contrary, our registration-fusion-denoising pipeline has better generalization and can be easily extended to the related image restoration tasks without major modifications. 
\section{Preliminary}\label{sec-preliminary}

\textbf{InstaHide.}
Given two private images $x_1$, $x_2$ and their corresponding one-hot labels $y_1$, $y_2$, InstaHide mixes $x_1$ and $x_2$ with $k-2$ public images to get a mixup image, and randomly flips the pixel signs of this mixup image to obtain the final encryption $m$, i.e.,
\begin{equation}\label{eq-instahide}
  m=\sigma \circ (\lambda_1x_1 + \lambda_2x_2 + \sum_{i=3}^{k} \lambda_iu_i),
\end{equation}
where $\lambda_i$ $(i\in \{1,\cdots,k\})$ is randomly sampled from $[0, 1]$ such that $\sum_{i=1}^{k}\lambda_i=1$, and all the images are normalized into $[-1, 1]$ beforehand. $\sigma$ is a one-time pad mask uniformly sampled from $\{+1, -1\}$, and $\circ$ denotes the element-wise multiplication. Accordingly, the label of $m$ becomes ${y}_m=\lambda_1y_1 + \lambda_2y_2$. The mixup pair $(m, {y}_m)$ is used to train the desired deep neural networks. 
%
%
Notice that the $k-2$ public images $u_i$ ($i\in \{3,\cdots, k\}$), randomly sampled from a public dataset (e.g., ImageNet~\cite{deng2009imagenet}), are mainly used to corrupt the visual features of $x_1$ and $x_2$, such that another party who obtains $m$ can not discern the original private images. As the public images are useless to the downstream classification tasks, we call the $\sum_{i=3}^{k} \lambda_iu_i$ term in Eq.~\ref{eq-instahide} the \textit{noise component}.

\begin{figure}[t]
\centering
\includegraphics[width=0.4\columnwidth]{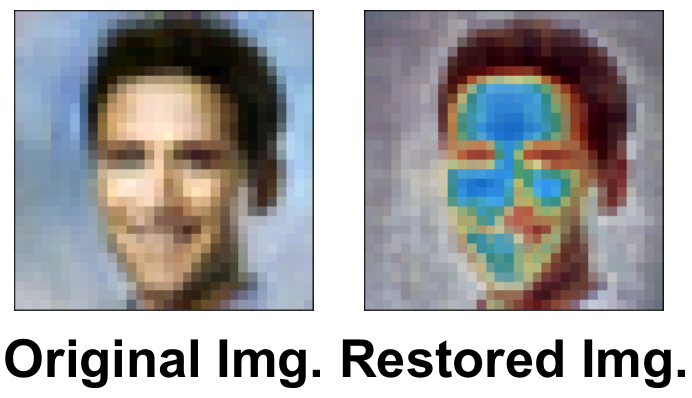}
\caption{Comparison between the original image (left) and the color shift image restored by \cite{carlini2020attack} (right).}
\label{fig-colorshift}
\end{figure}

\noindent \textbf{Carlini et al.'s Attack on InstaHide.}\label{sec-carlini}
Carlini et al.~\cite{carlini2020attack} propose an attack to restore the  private images $\mathcal{X}$ contained in the InstaHide Challenge dataset $\mathcal{M}$~\cite{instachallenge}.
The main idea is to first cluster the Challenge dataset, such that the encryptions in the same cluster contain the information of the same private image $x \in \mathcal{X}$. After that, for each cluster of encryptions, a gradient optimization method is employed to recover the private images by minimizing the $\ell_2$ norm of the noise component corresponding to the public images. Specifically, let a private image $x$ be a $d$-dimensional vector, $A$ be a $|\mathcal{X}|\times d$ private image matrix, $B$ be a $|\mathcal{M}|\times d$ encryption matrix and $C$ be a $|\mathcal{M}|\times |\mathcal{X}|$ coefficient matrix, i.e., each row of $A$ denotes a private image $x$, and each row of $B$ denotes an encryption image.
Therefore, Eq.~\ref{eq-instahide} can be rewritten as $\sigma \circ (C\cdot A + \delta) = B$, where $\delta$ denotes the noise component.  By preserving only the absolute pixel values, the randomness of pixel signs caused by $\sigma$ can be removed: $abs(C\cdot A +\delta) = abs(B)$.
Note that it is difficult to directly solve $A$ from this equation given $abs(B)$ and $C$ as the unknown noise component $\delta$ can significantly change the distribution of $C\cdot A$. Instead, \cite{carlini2020attack} proposes to solve a modified minimization problem:
\begin{equation}\label{eq-carlini}
  \argmin_{A'\in[-1,1]^{|\mathcal{X}|\times d}}  || \delta||^2_2 
  \text{\,\,\,\,s.t. \,}  C\cdot abs(A')+\delta = abs(B).
\end{equation}
%
However, there is a main defect in Eq.~\ref{eq-carlini} that $abs(C\cdot A)\neq C\cdot abs(A)$, for example, $abs\left(\begin{bmatrix}0.5 & 0.3\end{bmatrix}\cdot \begin{bmatrix}-0.8 & 1\end{bmatrix}^\intercal \right)=0.1$ whereas $\begin{bmatrix}0.5 & 0.3\end{bmatrix}\cdot abs\left(\begin{bmatrix}-0.8 & 1\end{bmatrix}^\intercal \right)=0.7$. Consequently, it will produce images with obvious color shifts, even leading to indistinguishable images (Fig.~\ref{fig-colorshift}).

\vspace{1mm}
\noindent \textbf{InstaHide with Data Augmentation.}
Similar to MixMatch~\cite{berthelot2019mixmatch},
let $\mathcal{X}=\{x_1, \cdots, x_N\}$ be a private image set, where $N$ is the number of private images. Before employing InstaHide, we conduct data augmentation on each image $x_i$ to generate a transformed dataset $\hat{\mathcal{X}}=\{\hat{x}_1,\cdots,\hat{x}_{N \times K}\}$. Specifically, we generate $K-1$ augmentations for each image $x_i$: 
$\hat{x}_{i,j}=\text{Augment}(x_i), j\in \{1, \cdots, K-1\}$. Meanwhile, we denote $x_i$ by $\hat{x}_{i,0}$. All $\hat{x}_{i,j} (i\in\{1,\cdots,N\} \wedge j \in \{0,\cdots,K-1\})$ are flattened into $\hat{\mathcal{X}}$.
After that, we shuffle $\hat{\mathcal{X}}$ to get $\hat{\mathcal{S}}=\{\hat{s}_1,\cdots, \hat{s}_{N \times K}\}$ and apply InstaHide on $\hat{\mathcal{X}}$ and $\hat{\mathcal{S}}$ to generate the encryption dataset $\mathcal{M} = \{m_1, \cdots, m_{N \times K}\}$ with
\begin{align}\label{eq-dataset}
\begin{split}
  m_i = \text{ InstaHide}(\hat{x}_i, \hat{s}_i, u_{i,3},\cdots,u_{i,k}), \forall i \in \{1,\cdots, N \times K\},
\end{split}
\end{align}
where $\{u_{i,3},\cdots,u_{i,k}\}$ are $k-2$ random public images, and $k$ is the mixup parameter in InstaHide. Accordingly, the labels $\mathcal{Y}_{\mathcal{M}} = \{y_1, \cdots, y_{N \times K}\}$ of $\mathcal{M}$ can be obtained:
  $y_i = \lambda_1^{}y_{\hat{x}_i} + \lambda_2^{}y_{\hat{s}_i}$, 
where $y_{\hat{x}_i}$ and $y_{\hat{s}_i}$ denote the one-hot labels of $\hat{x}_i$ and $\hat{s}_i$; $\lambda_1^{}$ and $\lambda_2^{}$ are the corresponding random coefficients. 
Consequently, $\mathcal{M}$ and $\mathcal{Y}_{\mathcal{M}}$ are used as the training dataset for classification tasks.
%

For data augmentation, we consider the \textit{geometric transformations}, e.g., random cropping, rotation, and translation, instead of noise injection and color transformation. Specifically, the former method is widely adopted in deep learning~\cite{OoiTWWCCGLTWXZZ15, shorten2019survey} (also included in the code of InstaHide~\cite{instahidecode}), and it can change the structures of images, bringing more difficulty to the restoration work since we have to align the structures of multiple transformed images before restoring the original one. In contrast, the effects of the latter methods are trivial and can be generally covered by the mixup noise introduced by InstaHide.
Note that the GAN‑based augmentation methods, which typically synthesize new images that are not included in the original private dataset, can be regarded as an upstream task of geometric transformations~\cite{shorten2019survey}. 
To better investigate the impact of geometric transformations on the security of InstaHide,
we formally define the augmentation level $\epsilon$ based on the pixel displacement as follows:
\begin{definition}[$\epsilon$-augmentation]
  Given an image $x$ with size $W\times H$ and its augmented version $\hat{x}$.
  Assume a pixel $p^{x}$ in $x$ has coordinate $C_{p^{x}}=(w_{p^{x}},h_{p^{x}})$; and a pixel $p^{\hat{x}}$ in $\hat{x}$ has coordinate $C_{p^{\hat{x}}}=(w_{p^{\hat{x}}} ,h_{p^{\hat{x}}})$. Then $\hat{x}$ is an $\epsilon$-augmentation of $x$ if for any possible pixel pairs $(p^{\hat{x}}, p^{x})$ that $p^{\hat{x}}$ is transformed from $p^{x}$, $w_{p^{\hat{x}}}\in [w_{p^{x}}-\frac{\epsilon}{2}W, w_{p^{x}}+\frac{\epsilon}{2}W]$ and $h_{p^{\hat{x}}}\in [h_{p^{x}}-\frac{\epsilon}{2}H, h_{p^{x}}+\frac{\epsilon}{2}H]$ hold.
\end{definition}
Fig.~\ref{fig-mixegg} shows an example of images with different data augmentation levels. Generally, the higher an $\epsilon$ is, the larger degree an image will be transformed with.
For example, $\epsilon=0.5$ corresponds to shifting $x$ left for $W/4$, or cropping $x$ to $3W/4\times 3H/4$.

%

\section{The Proposed Attack on InstaHide with Data Augmentation}\label{sec-attack-detail}
\begin{figure*}[t]
\centering
\includegraphics[width=1\textwidth]{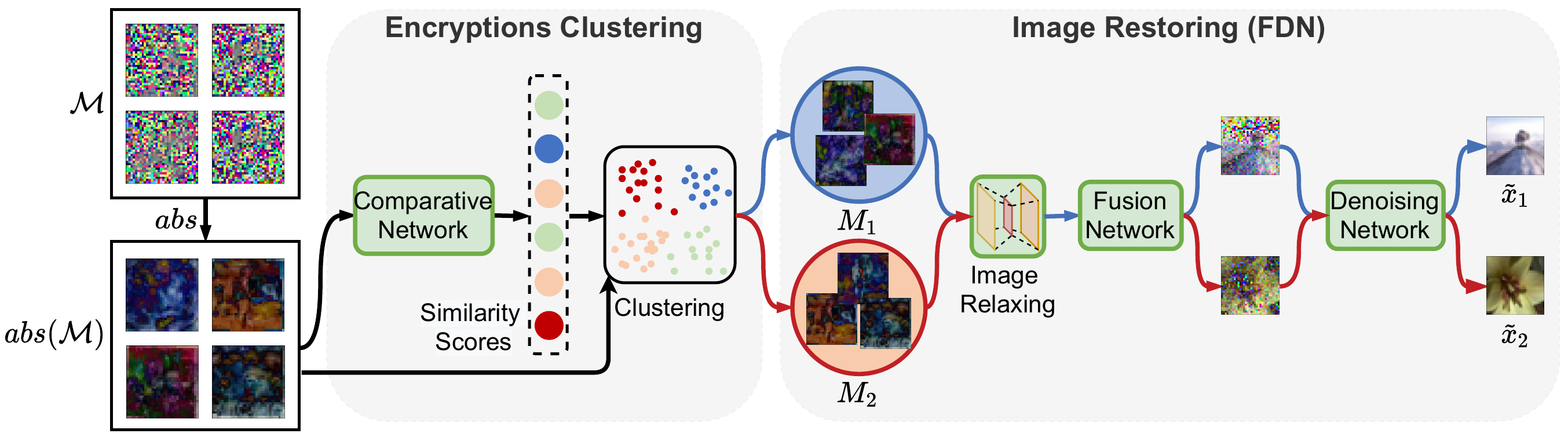}
\caption{Overview of the proposed attack.}
\label{fig-overview}
\end{figure*}

We consider a threat model in which the attacker aims to restore the data owner's private images $\mathcal{X}$ based on the published encryptions $\mathcal{M}$ and labels $\mathcal{Y}_{\mathcal{M}}$. We assume that the $(\mathcal{M}, \mathcal{Y}_{\mathcal{M}})$ is accessible to the attacker since \cite{huang2020instahide} claims that a data owner can directly send these data to another party for training desired models. 
Based on $\mathcal{M}$ and $\mathcal{Y}_{\mathcal{M}}$, our attack consists of three steps (Fig.~\ref{fig-overview}). In the \emph{absolute pre-processing} step, we remove the mask $\sigma$ (see Eq.~\ref{eq-instahide}) by conducting $abs(m_i)$, $\forall m_i \in \mathcal{M}$. $\sigma$  renders the  signs of mixup pixels to useless noise, yet failing to change the absolute pixel values. These absolute values can be utilized by the restoring algorithms. In the \emph{encryptions clustering} step, we find a candidate set of encryptions $M$ from $\mathcal{M}$ containing the information of the same private image $x_i$. In the \emph{image restoring} step, we use a fusion-denoising network (FDN) to restore ${x}_i$ from the homogeneous encryption set $M$. 
%




\subsection{Pixel-Level Noise Pattern}\label{subsec-noise-pattern}
InstaHide is a pixel-wise mixup scheme: the pixels located in the same position of different images are linearly combined into a mixup pixel, and the sign of this mixup pixel is randomly flipped. Formally, a pixel $p^{m_{i,l}}$ of an encrypted image $m_{i,l}$ is computed by: $p^{m_{i,l}} = \sigma \circ (\lambda_1p^{\hat{x}_i}+\lambda_2p^{\hat{s}_i}+p^{\delta})$,
where $p^{\hat{x}_i}$ and $p^{\hat{s}_i}$ are pixels from private images $\hat{x}_i$ and $\hat{s}_i$.
Without loss of generality, we assume $\hat{x}_i$ as the \emph{target image} that are commonly shared among a homogeneous encryption set, and the other image $\hat{s}_i$ as another source of noise. 
After conducting $abs(m_{i,l})$, the above equation can be rewritten as: $ abs(p^{m_{i,l}}) = abs(\lambda_1p^{\hat{x}_i}+p^{\delta})$.
Now the task becomes given $p^{m_{i,l}}$ and $\lambda_1$ (can be inferred from the one-hot labels), the adversary aims to restore the value of $p^{\hat{x}_i}$. This task is difficult if given only one encryption because  $p^{\delta}$ and the sign of $(\lambda_1p^{\hat{x}_i}+p^{\delta})$ are both unknown. 
We assume that the noise component $p^{\delta}$ follows a type of isotropic Gaussian distribution, which is reasonable since $p^{\delta}$ is initially a linear combination of $k-1$ images. 
Consequently, based on multiple $abs(p^{m_{i,l}})$ derived from a same $p^{\hat{x}_i}$, we can first roughly infer the signs of the corresponding $(\lambda_1p^{\hat{x}_i}+p^{\delta})$ by a neural network and then factor out the noise $p^{\delta}$ by averaging these $abs(p^{m_{i,l}})$ for restoring the original  $p^{\hat{x}_i}$.

The problem is how to find those $abs(p^{m_{i,l}})$ derived from a same $p^{\hat{x}_i}$. This is rather challenging in InstaHide with data augmentation, because the locations of corresponding $abs(p^{m_{i,l}})$ are mostly different from the original location of $p^{\hat{x}_i}$ after geometric transformations, and the visual features of the encryptions are barely useful for determining these transformations.
We observe that the neighboring pixels in an image patch typically change smoothly, i.e., their values are roughly the same, which means that the neighboring pixels of $\hat{x}_i$ can be used to align and recover $\hat{x}_i$. We therefore design an image-relaxing structure in the fusion phase to automatically diffuse the information of neighboring pixels into overlapping patches (information alignment), then use a window-based loss function in the denoising phase to patch-wisely restore the original image.





\subsection{Clustering Mixup Images}\label{subsec-com-nn}
To restore a private image $x_i$, we need to find a possible encryption set $M=\{m_{i,l}, l\in \{0,\cdots,a\}\}$ containing $x_i$ or its transformed versions $\hat{x}_{i,j}$ where $j \in \{1,\cdots,K-1\}$.
In this phase, we follow the clustering step  in~\cite{carlini2020attack}, i.e., splitting the encryptions $\mathcal{M}$ into multiple clusters such that the encryptions in each cluster contain the information of the same image. 
Note that the idea, i.e., first clustering the encryptions then recovering the corresponding private images, is inescapable for attacking InstaHide, because it is impossible to recover the original private image from only a single encryption given the random and severe corruptness \cite{huang2020instahide,carlini2020attack}.

\cite{carlini2020attack} uses a ResNet-28 to compute the similarity score for each pair of encryptions, which performs poorly in InstaHide with data augmentation. Because the periphery pixels produced by data augmentation are mostly useless for similarity comparison (Fig.~\ref{fig-mixegg}), yet severely degrading the comparison performance of ResNet-28 since it tries to remember all the peripheral patterns of training data.
Therefore, we design a new comparative network for computing the similarity scores (Fig.~\ref{fig-detail-comnet}).
Specifically, the multi-resolution information, which has been demonstrated beneficial in image comparison tasks~\cite{zagoruyko2015compare}, is used to help the network  pay more attention to the central pixels than the periphery pixels. For a $32\times 32$ image, we generate two $16\times 16$ images with different resolutions: the first image is generated by cropping the central part of the original image (high resolution), and the second image is generated by downsampling at half the original image (low resolution). For each pair of encryptions, we first 
generate a high-resolution pair and a low-resolution pair, then feed them into residual blocks~\cite{he2016deep}. The results are concatenated and fed into a dense layer for computing the final similarity score. 
As a result, in our experiments, the testing accuracy of the proposed network can reach 92$\%$ under $\epsilon=0.2$, whereas the accuracy of the original ResNet-28 reaches up to 71$\%$. 

\vspace{1mm}
\noindent\textbf{Additional Filtering.} After clustering, we obtain $|\mathcal{X}|$ clusters and each cluster consists of $|M|$ homogeneous encryptions, where $M=\{m_{i,l}, l\in \{0,\cdots,a\}\}$. 
In the experiments, we find that the encryptions with a large $\epsilon$ (e.g., rotated for 90 degrees), contribute little to or even degrade the restoration performance. This is because the structures of private images in these encryptions are difficult to be aligned with structures of other private images. Thus, we propose an additional filtering step to retain the neighboring encryptions in $M$ such that the $\epsilon$ difference between any two neighboring encryptions is less than a threshold $t_\epsilon$ with a high probability. Specifically, we train a filtering model based on an encryption dataset, where the encryption pairs with $\epsilon$ difference less than $t_\epsilon$ are labeled with $1$, and otherwise are labeled with $-1$. For each cluster $M$, we first use this filtering model to find all neighboring encryptions for each $m_{i,l}\in M$, then only keep the encryption $m$ with most neighbors (together with its neighbors) in $M$. As a result, we can guarantee that the $\epsilon$ difference of any two encryptions in $M$ is less than $2t_\epsilon$ as each encryption differs from $m$ by at most $t_\epsilon$. We conduct experiments with different $t_\epsilon$ and find that the filter with $t_\epsilon=0.2$ achieves a good trade-off between more homogeneous encryptions and less transformation after filtering $M$.


\subsection{Restoring Private Images}\label{subsec-fdn}
After obtaining a homogeneous encryption set $M$, we first use a re-weighting method to pre-process each encryption $m_{i,l}\in M$, then feed them into a fusion-denoising network to recover the target $\hat{x}_i$.

\vspace{1mm}
\noindent \textbf{Re-weighting.}
Notice that the coefficients $\lambda_1$ of $\hat{x}_i$ are different in different encryptions. The randomness of $\lambda_1$ may restrain the network from learning the correct pixel values of $\hat{x}_i$.
To reduce the uncertainty introduced by $\lambda_1$,
we rescale all encryptions by $1/\lambda_1$, i.e., for the pixels $p^{m_{i,l}}$ of $m_{i,l}$, we compute $abs(p^{m_{i,l}}/\lambda_1)=abs(p^{\hat{x}_i}+(p^{\delta}/\lambda_1))$.
%
Note that after rescaling, the corruptness levels of $\hat{x}_i$ are different in different encryptions.
For example, when $\lambda_1=0.4$ or 0.25,
the noise $p^{\delta}$ will be enlarged by a factor of 2.5 or 4, respectively. 
We further observe that the noise level $p^{\delta}/\lambda_1$ can be reflected by the variance of an encryption: assume $p^{\hat{x}_i}$ and $p^{\delta}$ are independent, then $\text{Var}(p^{\hat{x}_i}+\frac{p^{\delta}}{\lambda_1})=\text{Var}(p^{\hat{x}_i})+\frac{\text{Var}(p^{\delta})}{\lambda^2_1}$, which indicates that the larger $\lambda_1$ is, the smaller the variance of the encryption will be.
Based on this observation, we further re-weight the encryptions based on their variances. Specifically, we compute the variances $\text{Var}(m_{i,l})$ for each $m_{i,l}\in M$, then re-weight $m_{i,l}$ by a factor of $\beta = \frac{\min(Var(m_{i,0}),\cdots,Var(m_{i,a}))}{Var(m_{i,l})}$. The factor $\beta$ can ensure that the pixels of the encryption with the smallest variance (i.e., with the least corruptions) stay the same, while those with a larger variance are reduced since they contain more noise and provide less information of $\hat{x}_i$.
The effects of this re-weighting method are evaluated in the ablation study (Tab.~\ref{tab-ablation}).

\vspace{1mm}
\noindent \textbf{Image Fusion and Denoising.}
To accurately restore $\hat{x}_i$, we need to utilize all the information provided by each $m_{i,l}\in M$, i.e., fusing the information of these encryptions.
Recall that the $\hat{x}_i$ contained in an encryption $m_{i,l}$ could be either the original one or the transformed one.
Before fusion, we need to geometrically align these encryptions based on their respective target images~\cite{ma2019infrared}.
The traditional methods~\cite{rublee2011orb, ma2019infrared, lowe2004sift} are hardly useful since they rely on the visual features which are severely corrupted in this case (Fig.~\ref{fig-mixegg}).
Inspired by~\cite{dosovitskiy2016relax}, we design an efficient network component, called \emph{image relaxing}, to automatically align the information of target images.
%
%
%
Suppose the size of an encryption is $W\times H\times 3$. Before fusing the encryptions, we feed them into a convolutional layer with a stride of 2, resulting in a feature map with size $\lceil W/2\rceil\times \lceil H/2\rceil\times c$ (downsampling, $c$ is the number of filters). After that, we up-sample this feature map to the full image size $W\times H\times c$ by a transposed convolutional layer  with a stride of 2.
The downsampling step can produce translation-invariant features, and the upsampling step can capture the high-level structures.


\begin{figure}[t]
\centering
\includegraphics[width=.6\columnwidth]{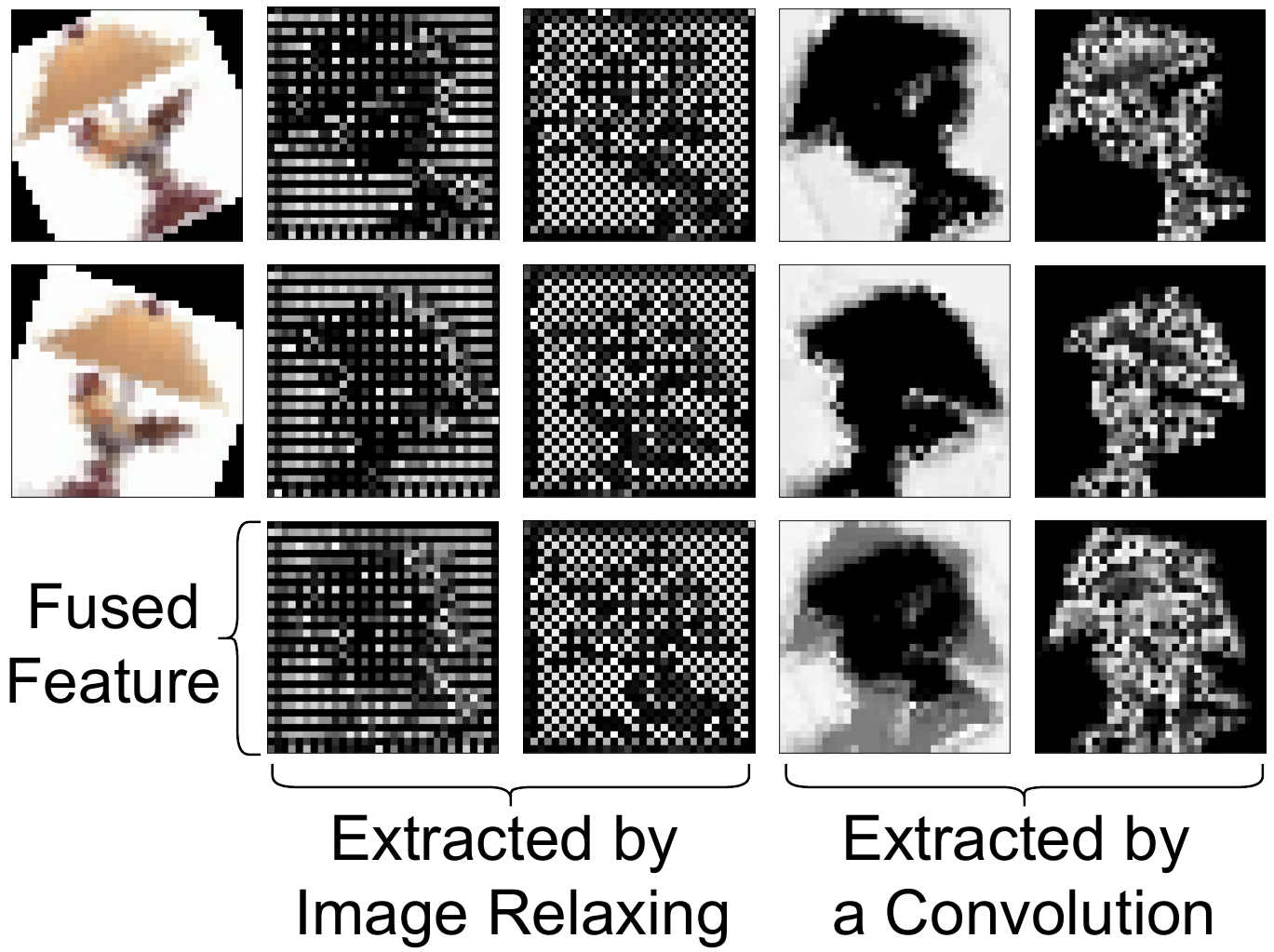}
\caption{A comparison between image relaxing and normal convolutions.}
\label{fig-imgrelax}
\end{figure}

To help illustrate the effects of image relaxing,
we first extract the features of two transformed images via a trained relaxing component and a $3\times 3$ convolutional kernel, then simply fuse the corresponding features by averaging them (Fig.~\ref{fig-imgrelax}). 
We observe that the features extracted by normal convolutions preserve more details, e.g., edges and corners, but the object structures are corrupted in the fused features because they are not properly aligned.
In the features extracted by image relaxing, some details are lost, but the original  structures experience less corruption after the fusion. Specifically, by downsampling and upsampling, image relaxing can transmit the information of a single pixel in the original image into a patch of neighboring pixels in the feature maps, which makes the information alignment easier, i.e., from point-wise alignment to patch-wise alignment. 
In addition, one encryption typically contains some structures irrelevant to the target private image. Image relaxing can lessen the impact of these irrelevant structures by downsampling, whereas normal convolutions preserve these structures and cause many artifacts on the fused images.

%

After extracting target features from multiple encryptions, we need to fuse these features based on a fusion rule. The two widely used rules are choose-max and average~\cite{ma2019infrared}.
In the experiments, we find that the choose-max rule performs better  when $|M|\leq 10$; while the average rule achieves better results when $|M|> 10$.
The reason is that the average rule could hardly factor out the noise $p^{\delta}$ based on a limited number (e.g., less than $10$) of encryptions, if some outliers, i.e., severely corrupted encryptions with large $p^{\delta}$, exist. While the choose-max rule mainly focuses on restoring the least corrupted $\hat{x}_i$ contained in the encryptions with larger pixel values (not reduced in the re-weighting phase), mitigating the impact of outliers.
Since the possible number $|M|$ of homogeneous encryptions input to \scheme\ could not be determined beforehand, we design a multiple-channel fusion architecture (Fig.~\ref{fig-detail-fdn}) to accept a variable number of encryptions as the input (all channels share the same set of parameters).
Specifically, for an input set $\{m_1,\cdots,m_n\}$, we first feed them into the re-weighting, relaxing and CovBlock components one-by-one (see Fig.~\ref{fig-detail-fdn}), and obtain a set of features $\{f(m_1),\cdots,f(m_n)\}$. These features will be merged into one feature $f(M)$ by the fusion component and fed into the denoising model.
Because the fusion phase can merge any number of inputs into one image for the following processing, \scheme\ can accept a variable number of encryptions as the input.

After fusing multiple encryptions, we use a denoising network to restore the original image.
Among multiple denoising networks (such as~\cite{mao2016rednet, zhang2020rdn, zhang2019rnan}),
we find RNAN~\cite{zhang2019rnan} performs best in this task since it can capture the long-range dependencies between channels and pixels in the whole image, which is important for \scheme\ to learn the overall noise distribution and restore private images with accurate color profiles.

\vspace{1mm}
\noindent \textbf{Loss Function.}
The mean structural similarity index (MSSIM)~\cite{wang2004mssim} performs far better than $\ell_1$ or $\ell_2$ loss in our task, since MSSIM compares the local difference between two images over a sliding window, which facilitates the network to neglect the overall structure distortion caused by other mixed images and concentrate on restoring local structures; whereas the other two losses tend to average all possible color modes and restore blurry images.
%
Since the $\ell_1$ loss can facilitate the recovery of pixel intensities~\cite{zhao2016loss}, we compute the network loss by combining the $\ell_1$ loss and MSSIM:
\begin{equation}\label{eq-loss}
  \mathcal{L}=\lambda_{\text{MSSIM}}\mathcal{L}_{\text{MSSIM}}+(1-\lambda_{\text{MSSIM}})\mathcal{L}_{\ell_1}.
\end{equation}

\section{Experiments}\label{sec-exp}

\begin{figure*}[t]
\begin{subfigure}[b]{.33\textwidth}
  \centering
  \includegraphics[width=1\textwidth]{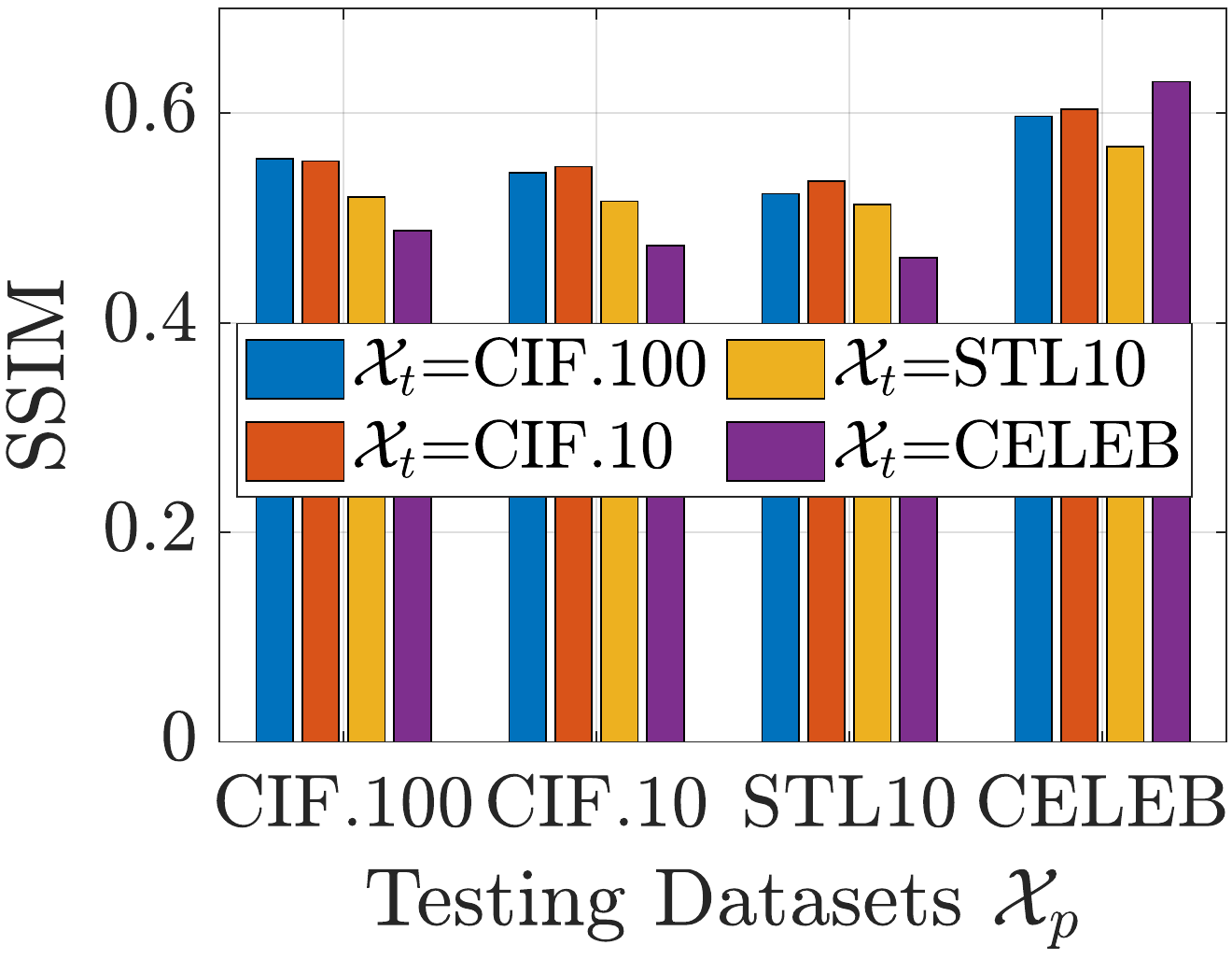}
  \caption{Private Datasets $\mathcal{X}$}
  \label{subfig-dataset-general}
\end{subfigure}
\begin{subfigure}[b]{.33\textwidth}
  \centering
  \includegraphics[width=1\textwidth]{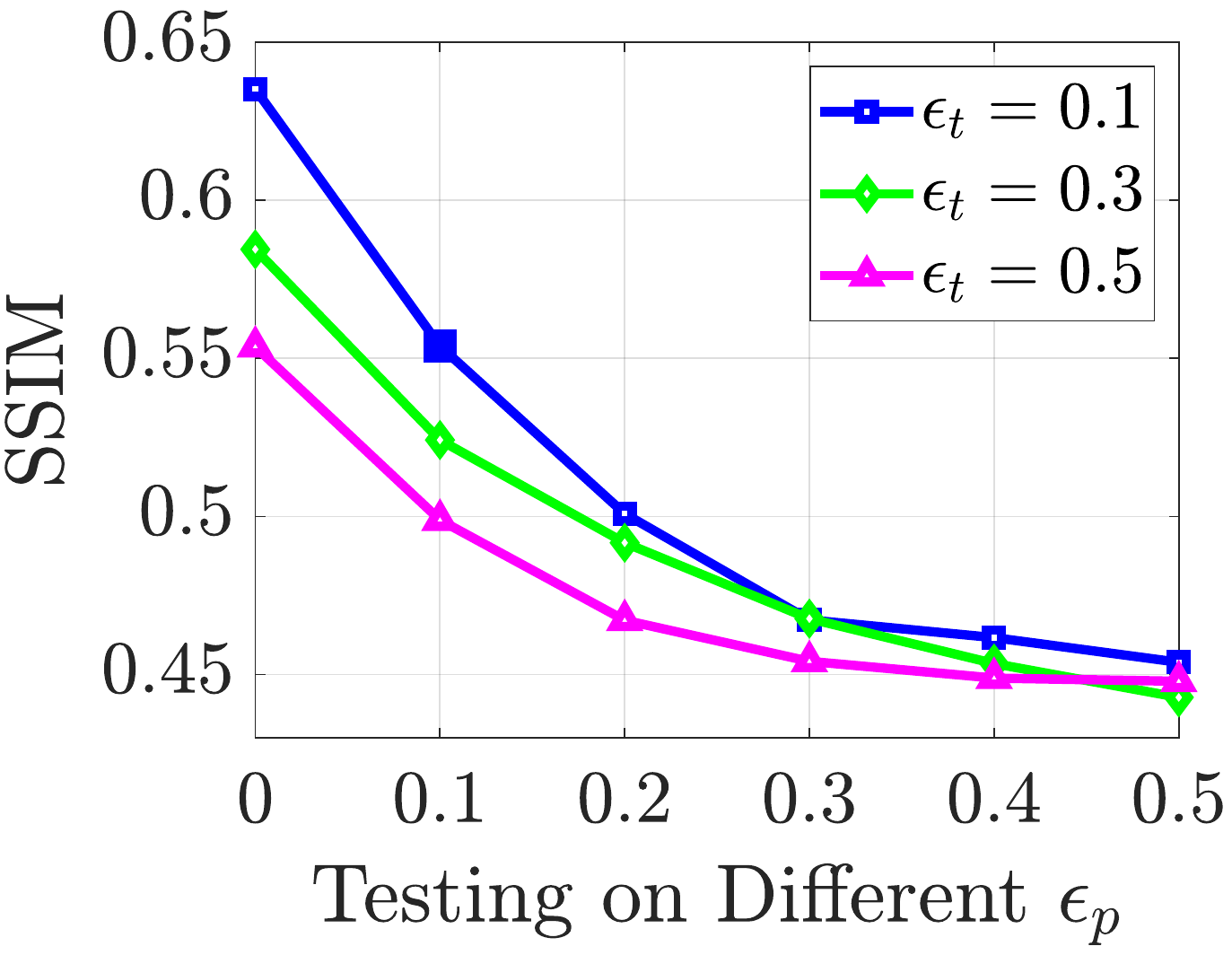}
  \caption{$\epsilon$-augmentation}
  \label{subfig-epsilon-general}
\end{subfigure}
\begin{subfigure}[b]{.33\textwidth}
  \centering
  \includegraphics[width=1\textwidth]{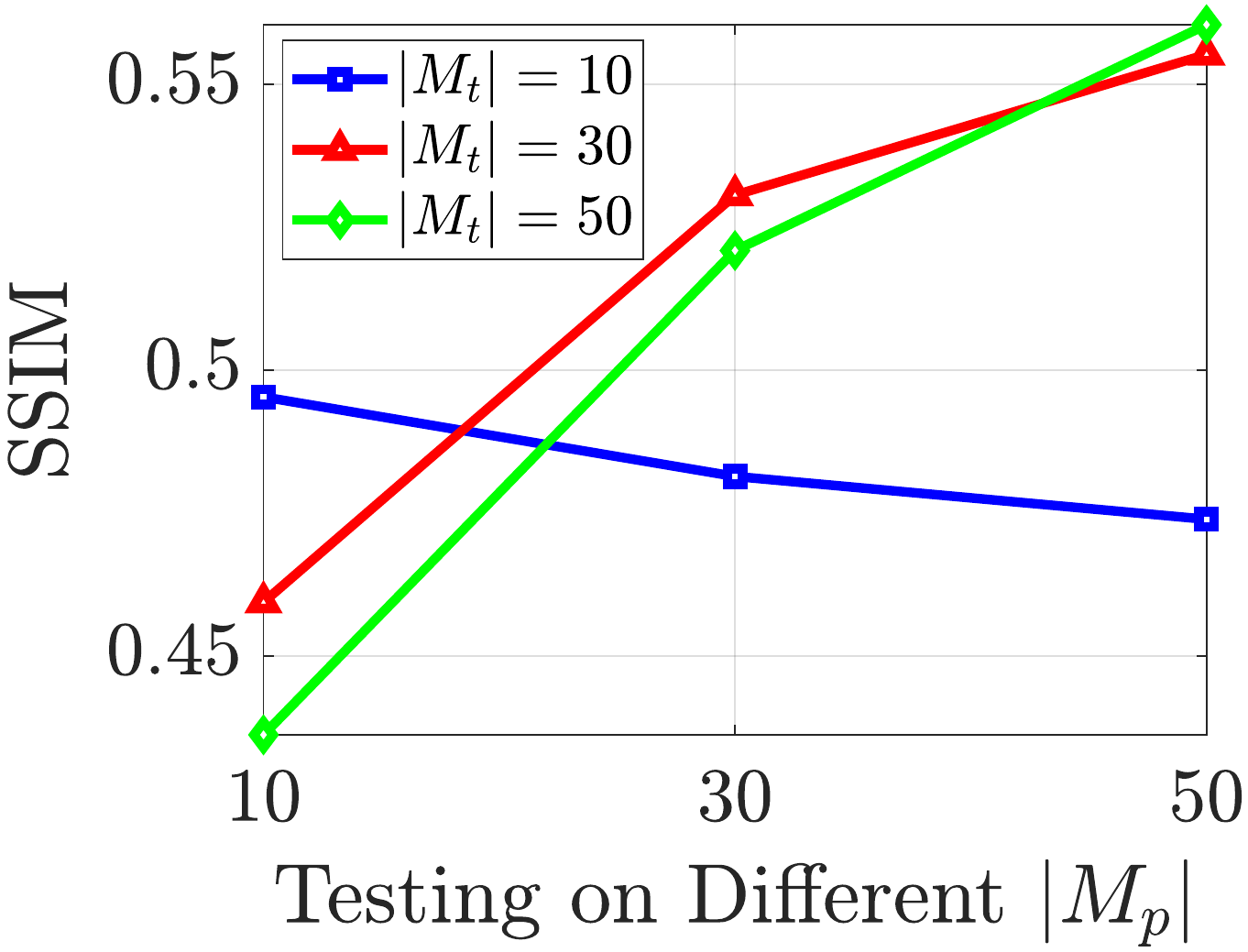}
  \caption{$|M|$}
  \label{subfig-M-general}
\end{subfigure}
\caption{The generalization performance of \scheme.}
\label{fig-general}
\end{figure*}

\textbf{Setup.}
We use CIFAR10~\cite{krizhevsky2009learning}, CIFAR100~\cite{krizhevsky2009learning}, STL10~\cite{coates2011analysis} and CelebFaces (CELEBA)~\cite{liu2015celeba} as the training and testing datasets. 
The $\lambda_{\text{MSSIM}}$ in Eq.~\ref{eq-loss} is empirically set to $0.7$. 
The InstaHide parameterized with $k=6$ is employed unless otherwise specified.
We compare our scheme with two baselines: the Carlini et al.'s original attack (CA)~\cite{carlini2020attack}, and a modified CA with the ResNet-28 used in the clustering phase replaced by our comparative network (CA-CN).
Besides, the MSSIM with window size 8 (SSIM for short) is used to measure the similarity between the restored images and the ground truth images. Note that the $\ell_1$ and $\ell_2$ (MSE) losses are not appropriate for the similarity evaluation in this paper, because they are pixel-wise metrics and a slight geometric transformation in the restored images could greatly change the results of them. More setting details are reported in the Appendix.

\vspace{-2mm}
\subsection{Generalization}\label{subsec-exp-generalization}
There are three hyper-parameters in FDN: the private image set $\mathcal{X}$ used for generating the training dataset of \scheme, the number $|M|$ of homogeneous encryptions in each cluster (i.e., the number of inputs to \scheme), and the data augmentation level $\epsilon$. We now demonstrate the generalization of \scheme\ with respect to the three hyper-parameters. Specifically, we investigate:
given a set of encryptions $\mathcal{M}_p$ which is generated from a private dataset ($\mathcal{X}_p$) with an unknown level of augmentation ($\epsilon_p$) and an unknown number of homogeneous encryptions derived from a same private image ($|M_p|$), whether the adversary can restore $\mathcal{X}_p$ via an \scheme\ trained on another encryption set $\mathcal{M}_t$ generated from different $\mathcal{X}_t$, $|M_t|$, and $\epsilon_t$. 

\vspace{1mm}
\noindent \textbf{Generalization w.r.t. different datasets $\mathcal{X}_p$.}
We fix $|M_p|=|M_t|=10$, $\epsilon_p=\epsilon_t=0.1$, and generate two encryption datasets $\mathcal{M}_p$ and $\mathcal{M}_t$ based on two  private datasets $\mathcal{X}_p$ and $\mathcal{X}_t$. We first train an \scheme\ based on $\mathcal{M}_t$ and then use it to  restore the $\mathcal{X}_p$ from $\mathcal{M}_p$. From Fig.~\ref{subfig-dataset-general}, we observe that the \scheme\ trained on CIFAR10 achieves the best restoration performance among the first three testing datasets, while the \scheme\ trained on CELEBA performs worst. Because the  image patterns in CIFAR10 are generally the most complicated among these datasets, which can help the network learn to restore images with complicated distributions;
whereas the image patterns of CELEBA (human faces) are more simple and predictable, making the networks trained on it perform worse on restoring more complicated images.
Note that when testing on the same dataset, the performance of the  \scheme s trained on different datasets remains roughly the same (suffering at most $6\%$ degradation), demonstrating \scheme's good generalization ability with respect to different datasets.


\vspace{1mm}
\noindent \textbf{Generalization w.r.t. different data augmentation levels $\epsilon_p$.}
We fix $|M_p|=|M_t|=10$ and $\mathcal{X}_p=\mathcal{X}_t=\text{CIFAR100}$ (80$\%$ for $\mathcal{X}_p$ and 20$\%$ for $\mathcal{X}_t$; this setting is used when $\mathcal{X}_p=\mathcal{X}_t$), then train and test an \scheme\ based on two encryption datasets generated with two different augmentation levels  $\epsilon_t$ and $\epsilon_p$. From Fig.~\ref{subfig-epsilon-general}, we see that with the increase of $\epsilon_p$, the performance of \scheme\ degrades. Because a larger $\epsilon_p$ represents a larger transformation to a private image, indicating that the structures of the private images contained in input encryptions are harder to be registered. 
Note that the \scheme\ trained on $\epsilon_t=0.1$ achieves better performance than other models.
The reason is that when trained on a dataset with smaller transformations, the \scheme\ learns to restore more image details instead of focusing on registering image structures.
Nevertheless, when tested on a specific $\epsilon_p$,
different \scheme s perform similarly, demonstrating the good generalization of \scheme\ under different augmentation levels.

\vspace{1mm}
\noindent \textbf{Generalization w.r.t. different number of inputs $|M_p|$.}
We first fix $\mathcal{X}_p = \mathcal{X}_t=\text{CIFAR100}$ and $\epsilon_p=\epsilon_t=0.2$, then train and test an \scheme\ under different $|M_p|$ and $|M_t|$. From Fig.~\ref{subfig-M-general}, we  see that with the increase of $|M_p|$, the testing performance of \scheme s trained on $|M_t|=30$ and $50$ improves; while the performance of the \scheme\ trained on $|M_t|=10$ slightly degrades.
This is because we use the choose-max and average rules to train \scheme s with $|M_t|\leq10$ and $|M_t|>10$, respectively. The choose-max rule is more robust under severely corrupted encryptions than the average rule, producing images with better quality when $|M|\leq 10$. While for $|M|> 10$, the average rule can learn more details than the choose-max rule. But when tested on the same $|M_p|$, the performances of \scheme\ trained on $|M_t|=30$ or 50 are similar. This shows the flexibility of \scheme, i.e., the adversary can train an \scheme\ based on different $|M_t|$ without worrying about the quality degradation of restored images.



\begin{table*}[htbp]
\center
\small
\begin{tabular}{c|c|ccccc||ccccc}
\hline
\multirow{2}{*}{Dataset} &\multirow{2}{*}{Attack} & \multicolumn{5}{c||}{Different $|M|$} & \multicolumn{5}{c}{Different $\epsilon$}\\
\cline{3-12}
  &   & 10  & 20 & 30 & 40 & 50 &
   0.1  & 0.2  & 0.3 & 0.4 & 0.5  \\
\hline 
\hline
\multirow{3}{*}{CIFAR100}   & CA & 0.3433&	0.3776&	0.3824&	0.3951&	0.4051 & 0.3917&	0.3304&	0.3150&	0.2986&	0.2781\\
                            & CA-CN & 0.4085&	0.4580&	0.4657&	0.4827&	0.4961  &	0.5062&	0.4416&	0.4041&	0.3939&	0.3552 \\
                            & \scheme & \textbf{0.5565}& \textbf{0.5936}& \textbf{0.6229}& \textbf{0.6327}& \textbf{0.6458} &	\textbf{0.6618}&	\textbf{0.6208}&	\textbf{0.6009}&	\textbf{0.5507}&	\textbf{0.5487} \\
                        
\hline
\multirow{3}{*}{CIFAR10}    & CA & 0.3831&	0.4165&	0.4168&	0.4203&	0.4335
 & 0.4167&	0.3503&	0.3259&	0.3133&	0.2877\\
                            & CA-CN & 0.4677	&0.4941&	0.5106&	0.5206&	0.5277 &	0.5073&	0.4515&	0.4219&	0.4087&	0.3674\\
                            & \scheme & \textbf{0.5490}&	\textbf{0.5844}&	\textbf{0.6133}&	\textbf{0.6289}&	\textbf{0.6404} & 	\textbf{0.6384}&	\textbf{0.6029}&	\textbf{0.5795}&	\textbf{0.5449}&	\textbf{0.5320}\\
                         
\hline
\multirow{3}{*}{STL10}      & CA & 0.4098&	0.4382&	0.4430&	0.4495&	0.4530
 & 0.4430&	0.3561&	0.3378&	0.3161&	0.2849\\
                            & CA-CN & 0.5057&	0.5312&	0.5449&	0.5465&	0.5636 &0.5636&	0.4902&	0.4441&	0.4303&	0.3646\\
                            & \scheme & \textbf{0.5130}&	\textbf{0.5622}&	\textbf{0.5923}&	\textbf{0.6091}&	\textbf{0.6307} & 	\textbf{0.6429}&	\textbf{0.5872}&	\textbf{0.5612}&	\textbf{0.5111}&	\textbf{0.4958}\\
\hline
\multirow{3}{*}{CELEBA}     & CA & 0.3775&	0.3872&	0.3897&	0.3980&	0.4039
 & 0.3981&	0.3290&	0.3111&	0.2997&	0.2793\\  
                            & CA-CN & 0.4593&	0.4843&	0.4954&	0.5018&	0.5066 	&0.5132	&0.4275&	0.4069	&0.3848	&0.3404\\
                            & \scheme & \textbf{0.6302}&	\textbf{0.6613}&	\textbf{0.6777}&	\textbf{0.6895}&	\textbf{0.7166} &	\textbf{0.7166}&	\textbf{0.7032}&	\textbf{0.6832}&	\textbf{0.6671}&	\textbf{0.6264}\\
\hline
\end{tabular}
\caption{The performance comparison (SSIM) \textit{w.r.t.} different number of inputs $|M|$ and different $\epsilon$-augmentation.}\label{tb-m-epsilon}
\end{table*}

\begin{figure*}[t]
\centering
\begin{subfigure}[b]{.45\textwidth}
  \centering
  \includegraphics[width=0.99\textwidth]{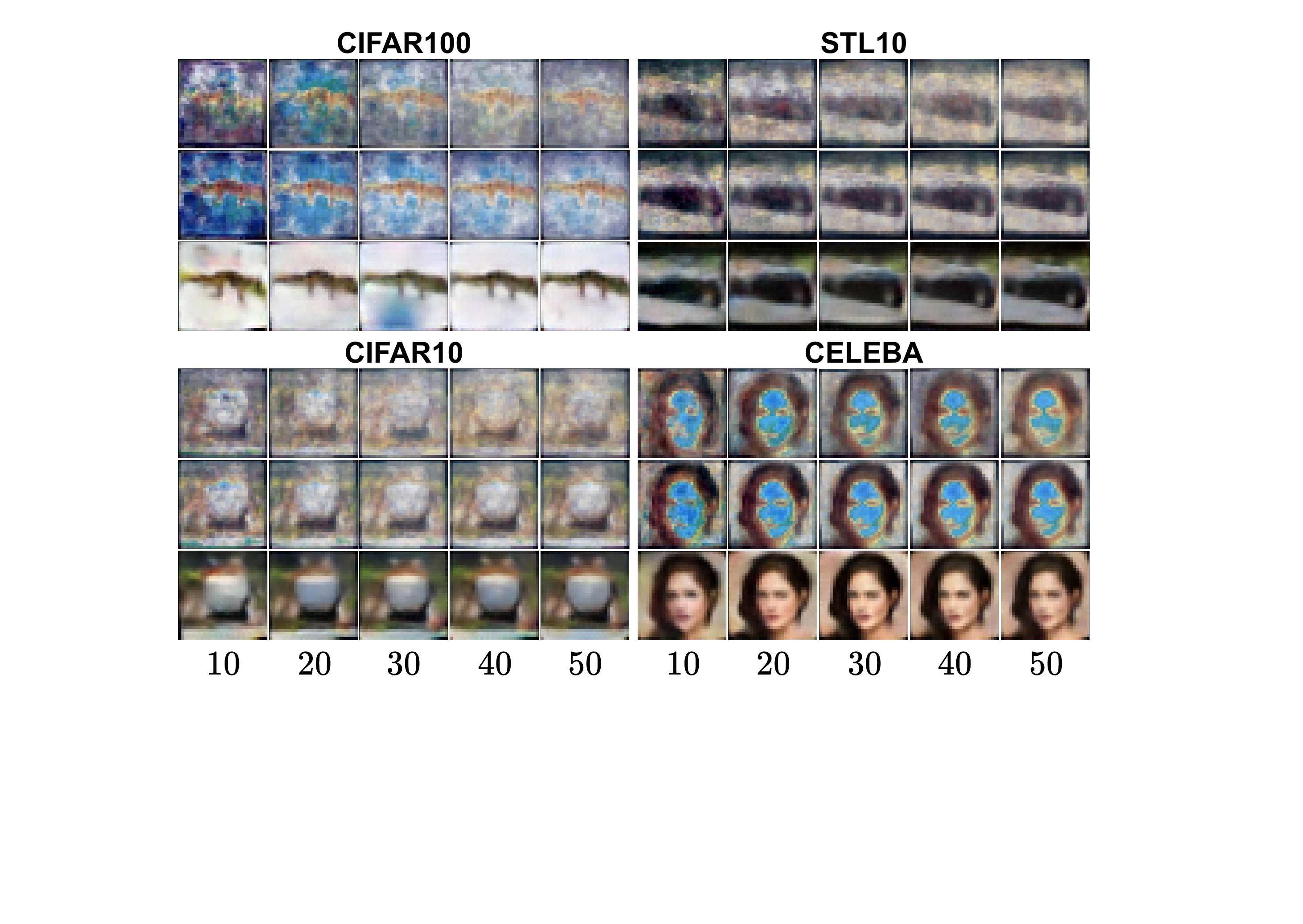}
  \caption{Examples \textit{w.r.t.} $|M|$}
  \label{fig-different-M}
\end{subfigure}
~
\begin{subfigure}[b]{.45\textwidth}
  \centering
  \includegraphics[width=0.99\textwidth]{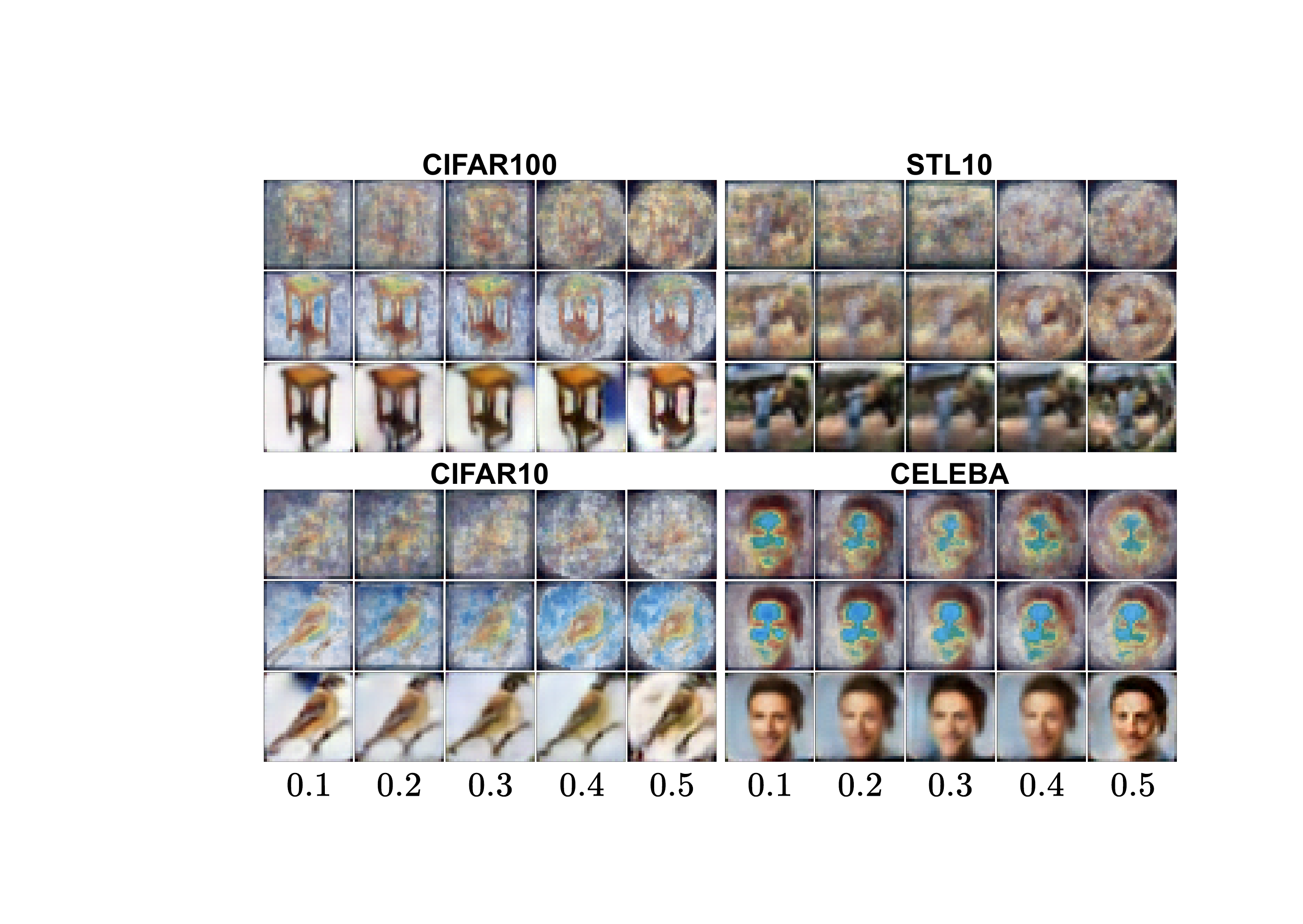}
  \caption{Examples \textit{w.r.t.} $\epsilon$}
  \label{fig-different-epsilon}
\end{subfigure}
\caption{The comparison of restored images \textit{w.r.t.} (a) different $|M|$  and (b) different $\epsilon$. The first row shows results of CA; the second row shows results of CA-CN; and the third row shows results of \scheme.  See Fig.~\ref{fig-differentM-fullexp} and \ref{fig-differentexp-fullexp} for more examples.}
\label{fig-compare-carlini}
\end{figure*}

\subsection{Comparison with Carlini et al.'s Attack}\label{subsec-exp-comparison}

\textbf{Different numbers of input mixups.}
In this set of experiments, we fix $\epsilon_p=\epsilon_t=0.1$, then train an \scheme\ based on $\mathcal{X}_t=\text{CIFAR10}$ and $|M_t|=30$, and test it under different $|M_p|$ (i.e., number of input encryptions to \scheme). We show the SSIM results in Tab.~\ref{tb-m-epsilon} and some restored images in Fig.~\ref{fig-different-M}.
From Tab.~\ref{tb-m-epsilon}, we see that the modified attack CA-CN performs better than the original attack CA. The reason is that the ResNet used in CA can produce plenty of false positive images (i.e., not containing the target private image) in the same cluster, which brings considerable noise to the input of the restoration phase and renders the final images indistinguishable. Our comparative network in CA-CN can reduce the false positive cases and transmit more useful information to the restoration algorithm.
In addition, with the increasing of $|M_p|$, both \scheme\ and CA can restore the private images with increasing quality. This is expected since more input encryptions can provide more details of the private images.
Also, \scheme\ performs better than CA, which can be clearly demonstrated by the examples in Fig.~\ref{fig-different-M}.
The substantial difference between the images restored by \scheme\ and images restored by CA is in the color profile.
\scheme\ can precisely restore the color profile and salient features, while CA loses considerable details and generates color shift areas in the restored images (as discussed in the preliminary section), which is most obvious in CELEBA.
%

%



\vspace{1mm}
\noindent \textbf{Different augmentation levels.}
We first fix $|M_p|=50$ and generate different testing encryption datasets from different $\mathcal{X}_p$ with different $\epsilon_p$, then attack these datasets via an \scheme\ trained on encryptions generated from $\mathcal{X}_t=\text{CIFAR10}$, $|M_t|=30$ and $\epsilon_t=0.1$.
Note that we use the \textit{filtering} phase  to process each set of homogeneous encryptions before inputting them to \scheme.
Tab.~\ref{tb-m-epsilon} and Fig.~\ref{fig-different-epsilon} show the restoration performance and some examples.
From Tab.~\ref{tb-m-epsilon}, we observe that the performance of \scheme\ degrades with the increasing of $\epsilon_p$. The main reason is that the filtering step reduces the number of homogeneous encryptions input to \scheme. Specifically, the general numbers of encryptions input to \scheme\ after filtering are 25, 19 and 16 corresponding to $\epsilon_p=0.3$, 0.4, 0.5, respectively. Less number of encryptions contain less information for the restoration of target images, causing the performance degradation of \scheme.
Nevertheless, Fig.~\ref{fig-different-epsilon} shows that  \scheme\ can restore far better colors and structures than CA.
Note that CA is a pixel-wise optimization method which is developed to restore the private images without any transformations. When recovering images pre-processed by data augmentation, the corresponding pixels from different encryptions could be unaligned, and are thus treated as noise and factored out by CA, leading to considerable detail loss.

\begin{table*}[t]
\centering
\begin{tabular}{  
>{\centering\arraybackslash} m{1.8cm}  
>{\centering\arraybackslash} m{1.3cm} >{\centering\arraybackslash} m{1.3cm}  >{\centering\arraybackslash} m{1.4cm} >{\centering\arraybackslash} m{1.3cm}  >{\centering\arraybackslash} m{1.3cm}  >{\centering\arraybackslash} m{1.3cm}
>{\centering\arraybackslash} m{1.5cm}   }
\toprule
Branch & Ground Truth & Original Network & No Re- weighting & No Img. Relaxing & $\ell_1$ loss & $\ell_2$ loss & Resnet Clustering \\
\toprule
Mean SSIM & / & 0.538 & 0.487 &  0.503 &  0.507 &  0.479 & 0.376 \\
An Example
& \includegraphics[width=0.06\textwidth]{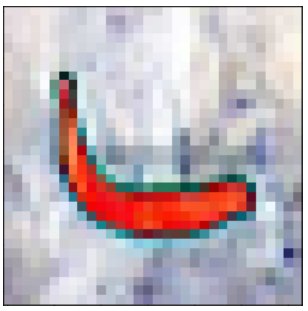} 
& \includegraphics[width=0.06\textwidth]{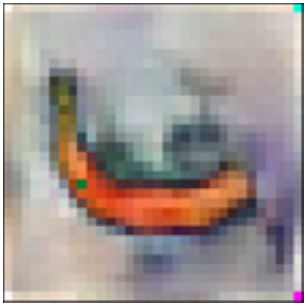} 
& \includegraphics[width=0.06\textwidth]{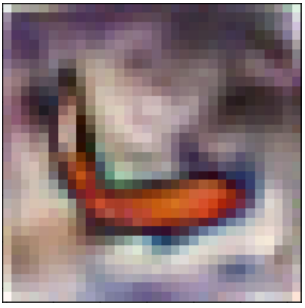} 
& \includegraphics[width=0.06\textwidth]{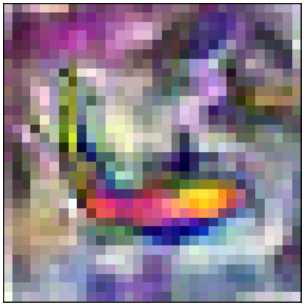} 
& \includegraphics[width=0.06\textwidth]{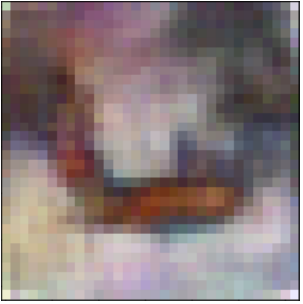} 
& \includegraphics[width=0.06\textwidth]{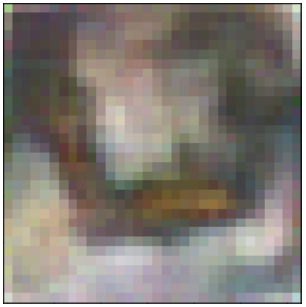} 
& \includegraphics[width=0.06\textwidth]{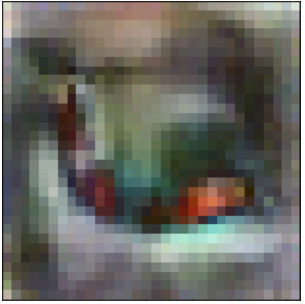}\\
\bottomrule
\end{tabular}
\caption{The results of ablation studies.\label{tab-ablation}}
\end{table*}

\subsection{Ablation}\label{exp-ablation}
We conduct ablation studies to validate the effectiveness of each component in the proposed attack. There are three key components in \scheme: image re-weighting, image relaxing, and the loss function.
We fix $|M_p|=|M_t|=10$, $\epsilon_p=\epsilon_t=0.1$ and $\mathcal{X}_p=\mathcal{X}_t=\text{CIFAR100}$, then remove each component from \scheme\ in turn and test the performance of the modified network. For image relaxing, we replace it with a $3\times3$ convolutional layer to extract the same number of features as performed by the original component. We replace the loss function with $\ell_1$ or $\ell_2$ and compare their results. In addition, we validate the effects of the comparative network by replacing it with a Resnet in the clustering phase.
The overall SSIM averaged on the testing images and some restored images are shown in Tab.~\ref{tab-ablation},
from which we observe that the \scheme\ without image re-weighting produces images with distorted pixels and noisy background. By image re-weighting, we reduce the impact of severely corrupted encryptions  on the quality of the restored images.
Image relaxing can automatically register the information of private images. The \scheme\ trained without image relaxing produces images with fuzzy boundaries and considerable artifacts.
In addition, $\ell_1$ loss can restore the original color profiles more accurately than $\ell_2$ loss. But they can hardly restore the image  structures. We also see that the original Resnet greatly degrades the restoration performance of \scheme, which is because it can not accurately compute the similarity between two encryptions, thus clustering many false positive encryptions (i.e., not containing the information of the target private image) into a homogeneous encryption set.

In addition, the results of attacking the InstaHide Challenge and the classification utility tests of InstaHide with different $\epsilon$ are reported in the Appendix. 
\section{Conclusion and Discussion}\label{sec-conclusion}

In this paper, we demonstrate that \cite{carlini2020attack} can not properly work under the real-world variations of InstaHide (decreasing the mixup times of a private image, or incorporating geometrically transformations before mixup). We accordingly design a fusion-denoising attack (\scheme) on the possible variations of InataHide to demonstrate that these variations can hardly reduce the privacy risks suffered by InstaHide. Although image relaxing  could cause some detail loss and reduce the sharpness of the restored images, the experiments demonstrate that \scheme\ can precisely restore the color profiles and structures, which issues an alert to the ML applications that seek to use a revised version of InstaHide to protect the private images. Nevertheless, the motivation of InstaHide, i.e., corrupting the visual features of private images, is promising in future studies. 
Some nonlinear transformations, instead of randomly flipping the pixel signs, can be further exploited for image encryption.
We believe more secure methods that incorporate data encryption into machine learning will play an important role in both the security community and AI systems.



\section*{Acknowledgments}
This research is supported by Singapore Ministry of Education Academic Research Fund Tier 3 under MOEs official
grant number MOE2017-T3-1-007.

\bibliography{references}

\begin{thebibliography}{43}
\providecommand{\natexlab}[1]{#1}

\bibitem[{ins(2020{\natexlab{a}})}]{instachallenge}
 2020{\natexlab{a}}.
\newblock A Challenge for InstaHide.
\newblock \url{https://github.com/Hazelsuko07/InstaHide_Challenge}.
\newblock Online; accessed 1-February-2021.

\bibitem[{ins(2020{\natexlab{b}})}]{instahidecode}
 2020{\natexlab{b}}.
\newblock InstaHide Training.
\newblock \url{https://github.com/Hazelsuko07/InstaHide}.
\newblock Online; accessed 15-May-2021.

\bibitem[{Arora(2020)}]{websanj}
Arora, S. 2020.
\newblock How to allow deep learning on your data without revealing the data.
\newblock \url{http://www.offconvex.org/2020/11/11/instahide/}.
\newblock Online; accessed 15-January-2021.

\bibitem[{Berthelot et~al.(2019)Berthelot, Carlini, Goodfellow, Papernot,
  Oliver, and Raffel}]{berthelot2019mixmatch}
Berthelot, D.; Carlini, N.; Goodfellow, I.; Papernot, N.; Oliver, A.; and
  Raffel, C.~A. 2019.
\newblock Mixmatch: A holistic approach to semi-supervised learning.
\newblock In \emph{{NeurIPS}}, 5049--5059.

\bibitem[{Carlini et~al.(2020)Carlini, Deng, Garg, Jha, Mahloujifar, Mahmoody,
  Song, Thakurta, and Tramer}]{carlini2020attack}
Carlini, N.; Deng, S.; Garg, S.; Jha, S.; Mahloujifar, S.; Mahmoody, M.; Song,
  S.; Thakurta, A.; and Tramer, F. 2020.
\newblock An Attack on InstaHide: Is Private Learning Possible with Instance
  Encoding?
\newblock \emph{arXiv preprint arXiv:2011.05315}.

\bibitem[{Coates, Ng, and Lee(2011)}]{coates2011analysis}
Coates, A.; Ng, A.; and Lee, H. 2011.
\newblock An analysis of single-layer networks in unsupervised feature
  learning.
\newblock In \emph{{AISTATS}}, 215--223.

\bibitem[{Deng et~al.(2009)Deng, Dong, Socher, Li, Li, and
  Fei-Fei}]{deng2009imagenet}
Deng, J.; Dong, W.; Socher, R.; Li, L.-J.; Li, K.; and Fei-Fei, L. 2009.
\newblock Imagenet: A large-scale hierarchical image database.
\newblock In \emph{2009 IEEE conference on computer vision and pattern
  recognition}, 248--255. Ieee.

\bibitem[{Dosovitskiy and Brox(2016)}]{dosovitskiy2016relax}
Dosovitskiy, A.; and Brox, T. 2016.
\newblock Inverting visual representations with convolutional networks.
\newblock In \emph{{CVPR}}, 4829--4837.

\bibitem[{Fredrikson, Jha, and Ristenpart(2015)}]{fredrikson2015model}
Fredrikson, M.; Jha, S.; and Ristenpart, T. 2015.
\newblock Model inversion attacks that exploit confidence information and basic
  countermeasures.
\newblock In \emph{{CCS}}, 1322--1333.

\bibitem[{Fu et~al.(2019)Fu, Wang, Xu, Mi, and Wang}]{fu2019mixup}
Fu, Y.; Wang, H.; Xu, K.; Mi, H.; and Wang, Y. 2019.
\newblock Mixup Based Privacy Preserving Mixed Collaboration Learning.
\newblock In \emph{{SOSE}}, 275--2755. IEEE.

\bibitem[{He et~al.(2016)He, Zhang, Ren, and Sun}]{he2016deep}
He, K.; Zhang, X.; Ren, S.; and Sun, J. 2016.
\newblock Deep residual learning for image recognition.
\newblock In \emph{Proceedings of the IEEE conference on computer vision and
  pattern recognition}, 770--778.

\bibitem[{Hitaj, Ateniese, and Perez-Cruz(2017)}]{hitaj2017deep}
Hitaj, B.; Ateniese, G.; and Perez-Cruz, F. 2017.
\newblock Deep models under the GAN: information leakage from collaborative
  deep learning.
\newblock In \emph{{CCS}}, 603--618.

\bibitem[{Huang et~al.(2020)Huang, Song, Li, and Arora}]{huang2020instahide}
Huang, Y.; Song, Z.; Li, K.; and Arora, S. 2020.
\newblock Instahide: Instance-hiding schemes for private distributed learning.
\newblock In \emph{{ICML}}, 4507--4518.

\bibitem[{Kairouz et~al.(2019)Kairouz, McMahan, Avent, Bellet, Bennis, Bhagoji,
  Bonawitz, Charles, Cormode, Cummings et~al.}]{kairouz2019advances}
Kairouz, P.; McMahan, H.~B.; Avent, B.; Bellet, A.; Bennis, M.; Bhagoji, A.~N.;
  Bonawitz, K.; Charles, Z.; Cormode, G.; Cummings, R.; et~al. 2019.
\newblock Advances and open problems in federated learning.
\newblock \emph{arXiv preprint arXiv:1912.04977}.

\bibitem[{Kingma and Ba(2014)}]{kingma2014adam}
Kingma, D.~P.; and Ba, J. 2014.
\newblock Adam: A method for stochastic optimization.
\newblock \emph{arXiv preprint arXiv:1412.6980}.

\bibitem[{Krizhevsky, Hinton et~al.(2009)}]{krizhevsky2009learning}
Krizhevsky, A.; Hinton, G.; et~al. 2009.
\newblock Learning multiple layers of features from tiny images.

\bibitem[{Li et~al.(2018)Li, He, Tao, Tang, and Wang}]{li2018joint}
Li, H.; He, X.; Tao, D.; Tang, Y.; and Wang, R. 2018.
\newblock Joint medical image fusion, denoising and enhancement via
  discriminative low-rank sparse dictionaries learning.
\newblock \emph{Pattern Recognition}, 79: 130--146.

\bibitem[{Li et~al.(2020)Li, Sahu, Talwalkar, and Smith}]{li2020federated}
Li, T.; Sahu, A.~K.; Talwalkar, A.; and Smith, V. 2020.
\newblock Federated learning: Challenges, methods, and future directions.
\newblock \emph{IEEE Signal Processing Magazine}, 37(3): 50--60.

\bibitem[{Liu, Xu, and Fang(2020)}]{liu2020infrared}
Liu, L.; Xu, L.; and Fang, H. 2020.
\newblock Infrared and visible image fusion and denoising via {$\ell_2$} -
  {$\ell_p$} norm minimization.
\newblock \emph{Signal Processing}, 172: 107546.

\bibitem[{Liu, Liu, and Wang(2015)}]{liu2015transform}
Liu, Y.; Liu, S.; and Wang, Z. 2015.
\newblock A general framework for image fusion based on multi-scale transform
  and sparse representation.
\newblock \emph{Information fusion}, 24: 147--164.

\bibitem[{Liu et~al.(2015)Liu, Luo, Wang, and Tang}]{liu2015celeba}
Liu, Z.; Luo, P.; Wang, X.; and Tang, X. 2015.
\newblock Deep Learning Face Attributes in the Wild.
\newblock In \emph{{ICCV}}, 3730--3738.

\bibitem[{Lowe(2004)}]{lowe2004sift}
Lowe, D.~G. 2004.
\newblock Distinctive image features from scale-invariant keypoints.
\newblock \emph{International journal of computer vision}, 60(2): 91--110.

\bibitem[{Luo et~al.(2021)Luo, Wu, Xiao, and Ooi}]{luo2021feature}
Luo, X.; Wu, Y.; Xiao, X.; and Ooi, B.~C. 2021.
\newblock Feature inference attack on model predictions in vertical federated
  learning.
\newblock In \emph{2021 IEEE 37th International Conference on Data Engineering
  (ICDE)}, 181--192. IEEE.

\bibitem[{Ma, Ma, and Li(2019)}]{ma2019infrared}
Ma, J.; Ma, Y.; and Li, C. 2019.
\newblock Infrared and visible image fusion methods and applications: A survey.
\newblock \emph{Information Fusion}, 45: 153--178.

\bibitem[{Mao, Shen, and Yang(2016)}]{mao2016rednet}
Mao, X.-J.; Shen, C.; and Yang, Y.-B. 2016.
\newblock Image restoration using very deep convolutional encoder-decoder
  networks with symmetric skip connections.
\newblock \emph{arXiv preprint arXiv:1603.09056}.

\bibitem[{Mei, Dong, and Huang(2019)}]{mei2019simultaneous}
Mei, J.-J.; Dong, Y.; and Huang, T.-Z. 2019.
\newblock Simultaneous image fusion and denoising by using fractional-order
  gradient information.
\newblock \emph{Journal of Computational and Applied Mathematics}, 351:
  212--227.

\bibitem[{Ooi et~al.(2015)Ooi, Tan, Wang, Wang, Cai, Chen, Gao, Luo, Tung,
  Wang, Xie, Zhang, and Zheng}]{OoiTWWCCGLTWXZZ15}
Ooi, B.~C.; Tan, K.; Wang, S.; Wang, W.; Cai, Q.; Chen, G.; Gao, J.; Luo, Z.;
  Tung, A. K.~H.; Wang, Y.; Xie, Z.; Zhang, M.; and Zheng, K. 2015.
\newblock {SINGA:} {A} Distributed Deep Learning Platform.
\newblock In \emph{Proceedings of the {ACM} International Conference on
  Multimedia}, 685--688.

\bibitem[{Raynal, Achanta, and Humbert(2020)}]{raynal2020image}
Raynal, M.; Achanta, R.; and Humbert, M. 2020.
\newblock Image Obfuscation for Privacy-Preserving Machine Learning.
\newblock \emph{arXiv preprint arXiv:2010.10139}.

\bibitem[{Rublee et~al.(2011)Rublee, Rabaud, Konolige, and
  Bradski}]{rublee2011orb}
Rublee, E.; Rabaud, V.; Konolige, K.; and Bradski, G. 2011.
\newblock ORB: An efficient alternative to SIFT or SURF.
\newblock In \emph{{ICCV}}, 2564--2571.

\bibitem[{Shorten and Khoshgoftaar(2019)}]{shorten2019survey}
Shorten, C.; and Khoshgoftaar, T.~M. 2019.
\newblock A survey on image data augmentation for deep learning.
\newblock \emph{Journal of Big Data}, 6(1): 1--48.

\bibitem[{Wang et~al.(2004)Wang, Bovik, Sheikh, and Simoncelli}]{wang2004mssim}
Wang, Z.; Bovik, A.~C.; Sheikh, H.~R.; and Simoncelli, E.~P. 2004.
\newblock Image quality assessment: from error visibility to structural
  similarity.
\newblock \emph{IEEE transactions on image processing}, 13(4): 600--612.

\bibitem[{Wu et~al.(2020)Wu, Cai, Xiao, Chen, and Ooi}]{WuCXCO20}
Wu, Y.; Cai, S.; Xiao, X.; Chen, G.; and Ooi, B.~C. 2020.
\newblock Privacy Preserving Vertical Federated Learning for Tree-based Models.
\newblock \emph{Proc. {VLDB} Endow.}, 13(11): 2090--2103.

\bibitem[{Yang et~al.(2019)Yang, Liu, Chen, and Tong}]{yang2019federated}
Yang, Q.; Liu, Y.; Chen, T.; and Tong, Y. 2019.
\newblock Federated machine learning: Concept and applications.
\newblock \emph{{TIST}}, 10(2): 1--19.

\bibitem[{Zagoruyko and Komodakis(2015)}]{zagoruyko2015compare}
Zagoruyko, S.; and Komodakis, N. 2015.
\newblock Learning to compare image patches via convolutional neural networks.
\newblock In \emph{{CVPR}}, 4353--4361.

\bibitem[{Zhang et~al.(2017)Zhang, Cisse, Dauphin, and
  Lopez-Paz}]{zhang2017mixup}
Zhang, H.; Cisse, M.; Dauphin, Y.~N.; and Lopez-Paz, D. 2017.
\newblock mixup: Beyond empirical risk minimization.
\newblock \emph{arXiv preprint arXiv:1710.09412}.

\bibitem[{Zhang and Luo(2021)}]{zhang2021exploiting}
Zhang, X.; and Luo, X. 2021.
\newblock Exploiting Defenses against GAN-Based Feature Inference Attacks in
  Federated Learning.
\newblock arXiv:2004.12571.

\bibitem[{Zhang, Bai, and Wang(2017)}]{zhang2017boundary}
Zhang, Y.; Bai, X.; and Wang, T. 2017.
\newblock Boundary finding based multi-focus image fusion through multi-scale
  morphological focus-measure.
\newblock \emph{Information fusion}, 35: 81--101.

\bibitem[{Zhang et~al.(2019)Zhang, Li, Li, Zhong, and Fu}]{zhang2019rnan}
Zhang, Y.; Li, K.; Li, K.; Zhong, B.; and Fu, Y. 2019.
\newblock Residual non-local attention networks for image restoration.
\newblock \emph{arXiv preprint arXiv:1903.10082}.

\bibitem[{Zhang et~al.(2020{\natexlab{a}})Zhang, Liu, Sun, Yan, Zhao, and
  Zhang}]{zhang2020ifcnn}
Zhang, Y.; Liu, Y.; Sun, P.; Yan, H.; Zhao, X.; and Zhang, L.
  2020{\natexlab{a}}.
\newblock IFCNN: A general image fusion framework based on convolutional neural
  network.
\newblock \emph{Information Fusion}, 54: 99--118.

\bibitem[{Zhang et~al.(2020{\natexlab{b}})Zhang, Tian, Kong, Zhong, and
  Fu}]{zhang2020rdn}
Zhang, Y.; Tian, Y.; Kong, Y.; Zhong, B.; and Fu, Y. 2020{\natexlab{b}}.
\newblock Residual dense network for image restoration.
\newblock \emph{IEEE Transactions on Pattern Analysis and Machine
  Intelligence}.

\bibitem[{Zhao et~al.(2016)Zhao, Gallo, Frosio, and Kautz}]{zhao2016loss}
Zhao, H.; Gallo, O.; Frosio, I.; and Kautz, J. 2016.
\newblock Loss functions for image restoration with neural networks.
\newblock \emph{IEEE Transactions on computational imaging}, 3(1): 47--57.

\bibitem[{Zhu and Han(2020)}]{zhu2020deep}
Zhu, L.; and Han, S. 2020.
\newblock Deep leakage from gradients.
\newblock In \emph{Federated Learning}, 17--31. Springer.

\bibitem[{Zitova and Flusser(2003)}]{zitova2003imgreg}
Zitova, B.; and Flusser, J. 2003.
\newblock Image registration methods: a survey.
\newblock \emph{Image and vision computing}, 21(11): 977--1000.

\end{thebibliography}

\clearpage
\appendix

\section{Appendix}

\subsection{Clustering Details}\label{app-cluster}
\cite{carlini2020attack} uses a ResNet-28 model to compute the similarity score for each pair of encryptions. Based on these scores, we create a weighted graph, where the vertices denote the encryptions and the weighted edges denote the similarities between encryptions. Then, $|\mathcal{X}|$ densely cliques are identified, where $|\mathcal{X}|$ is the number of private images, and $|\mathcal{X}|<<|\mathcal{M}|$. Finally, the similarities between each encryption and each clique are computed, then each encryption can be mapped to two private images (i.e., cliques) since one encryption is derived from two private images.

\subsection{Experiment Details}\label{app-exp-detail}

\textbf{Dataset Generation.}
We generate the training encryptions based on the method described in the preliminary section, where the InstaHide is implemented based on~\cite{huang2020instahide}. Specifically, the augmented versions of each private image $x_i\in \mathcal{X}$ are generated by random rotation, random affine, and random crop followed by resizing.
Additionally, the training datasets of \scheme\ consist of $(M, y_{M})$ pairs, where $M$ is a set of homogeneous encryptions, and $y_{M}$ is the ground truth image that should be restored by \scheme\ based on $M$.
Note that the encryption with the smallest variance is called the \emph{reference encryption}, in which the target image $\hat{x}_i$ is least corrupted compared to other encryptions. Consequently, we focus on restoring $\hat{x}_i$ contained in the reference encryption.
We choose the target image contained in the reference encryption of $M$ as the label image $y_{M}$, i.e., $y_{M} = \{\hat{x}_{i,l}, l=\argmin_l\text{Var}(m_{i,l})\}$. The $(M, y_{M})$ pairs are then used to train the fusion-denoising network.

\begin{figure*}[t]
\centering
\includegraphics[width=.75\textwidth]{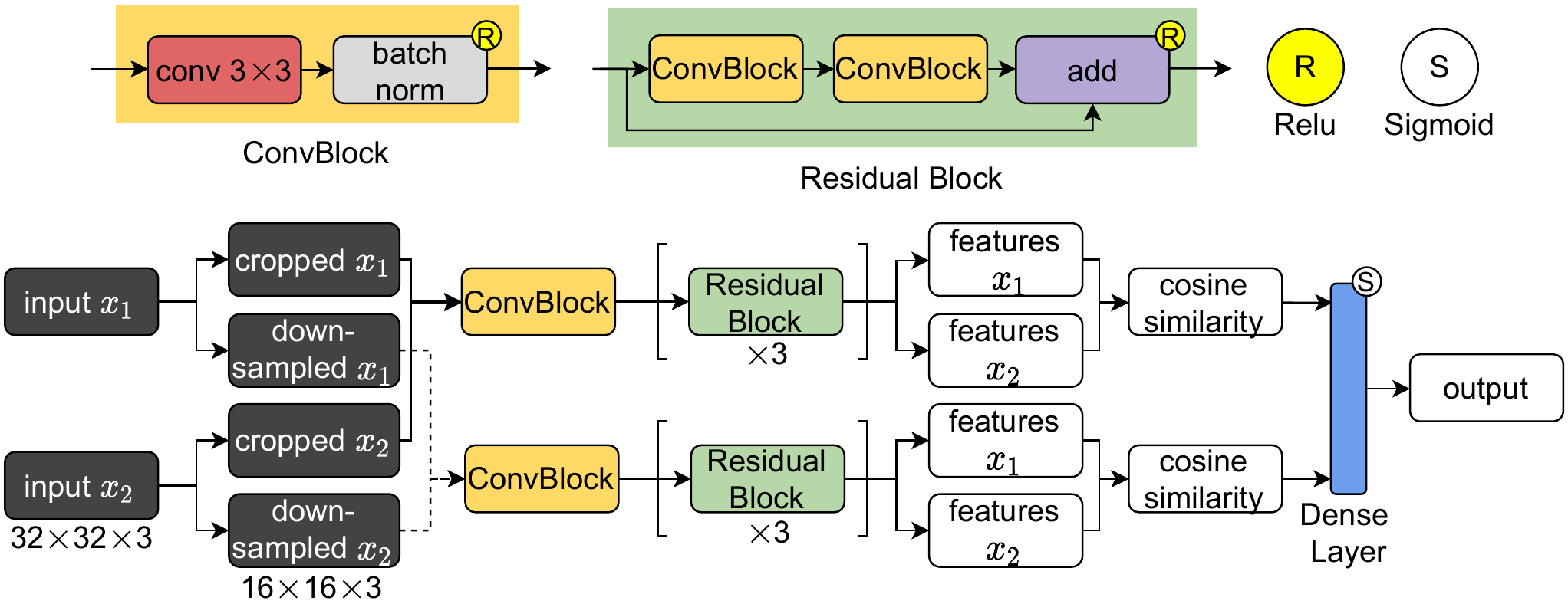}
\caption{The detailed architecture of the comparative neural network.}
\label{fig-detail-comnet}
\end{figure*}

\begin{figure*}[t]
\centering
\includegraphics[width=1\textwidth]{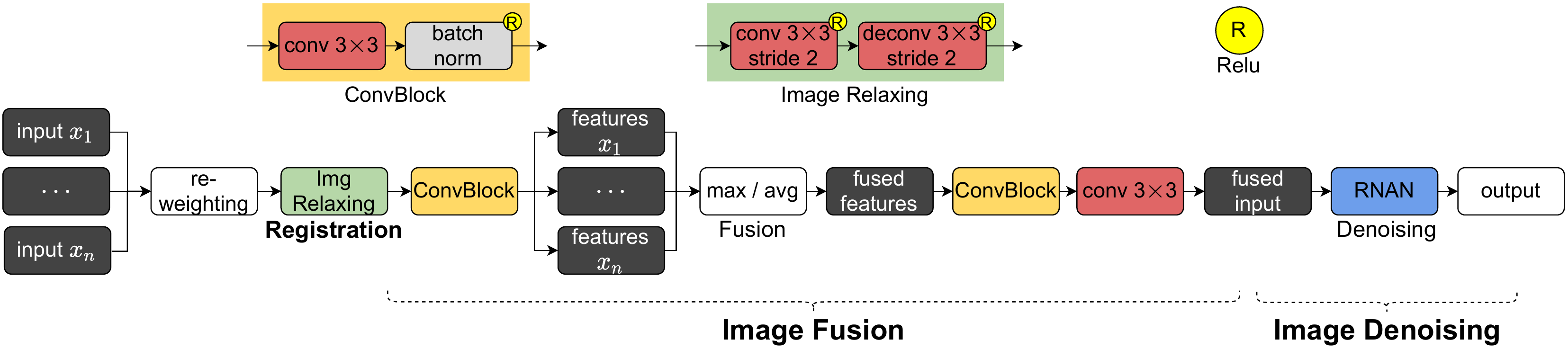}
\caption{The detailed architecture of the fusion-denoising network.}
\label{fig-detail-fdn}
\end{figure*}

\vspace{1mm}
\noindent \textbf{Experimental Setup.}
All the networks are implemented with PyTorch and trained with Adam optimizer~\cite{kingma2014adam} using an initial learning rate $10^{-4}$. \scheme\ is trained for 80 epochs, and the comparative network is trained for 100 epochs.
All the experiments are conducted on a platform with Intel Xeon Silver 4108 CPU and NVIDIA Tesla V100 GPU.
We select an RNAN with 6 residual blocks and 16 filters per block as the denoising component of \scheme. 
Training a comparative network and an \scheme\ takes about 3 h and 2 h, respectively. 
The detailed architectures of the comparative network and \scheme\ are shown in Fig.~\ref{fig-detail-comnet} and \ref{fig-detail-fdn}, respectively.

\emph{Datasets.} 
We use CIFAR10~\cite{krizhevsky2009learning}, CIFAR100~\cite{krizhevsky2009learning}, STL10~\cite{coates2011analysis} and CelebFaces Attributes  (CELEBA)~\cite{liu2015celeba} as the training and testing datasets. 
For simplicity, we first resize all images to 32$\times$32 and then generate encryptions based on the data generation method described in the preliminary section.

\emph{Baseline.} 
We compare our scheme with Carlini et al.'s scheme~\cite{carlini2020attack} with different settings. Since the ResNet-28 performs poorly against the InstaHide with data augmentation, we replace the ResNet used in the clustering phase of~\cite{carlini2020attack} with our comparative network and focus on comparing the restoration performance between \scheme\ and~\cite{carlini2020attack}. Besides, the MSSIM with window size 8 (SSIM for short) is used to measure the similarity between the restored images and the ground truth images. Note that the $\ell_1$ and $\ell_2$ (MSE) losses are not appropriate for the similarity evaluation in this paper, because they  perform poorly in evaluating perceptual image quality~\cite{wang2004mssim}, and a slight position shift in the restored images could lead to greatly different results of these two metrics.

\subsection{Additional Experiment Results}\label{app-exp}

\begin{figure*}[t]
\centering
\includegraphics[width=.8\columnwidth]{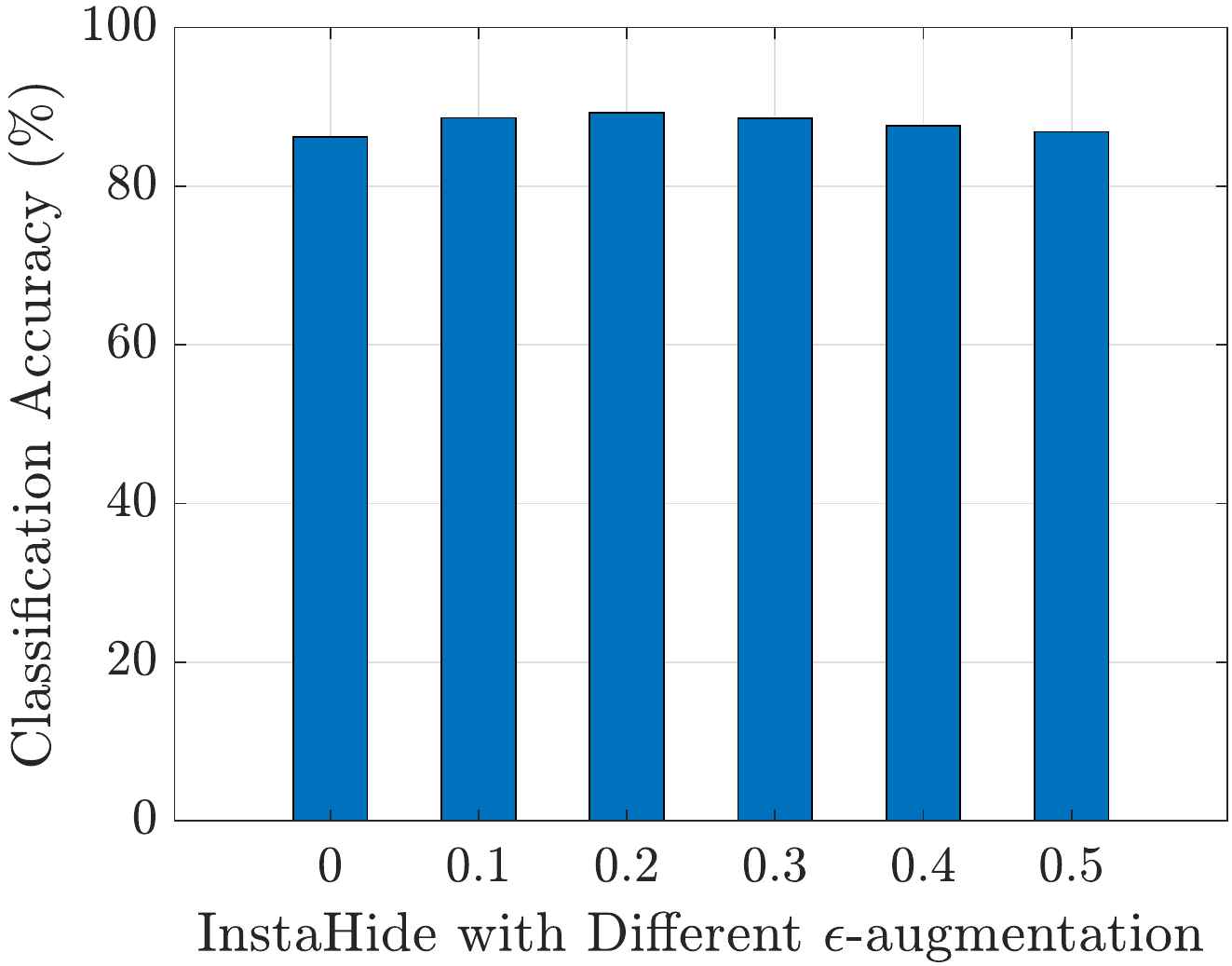}
\caption{The testing accuracies of ResNet18 trained on InstaHide with different $\epsilon$ (CIFAR10).}
\label{subfig-insta-accur}
\end{figure*}

\noindent \textbf{Classification utility of InstaHide with data augmentation.}
We validate the classification utility of InstaHide with different $\epsilon$, as shown in Fig.~\ref{subfig-insta-accur}. We observe that the testing accuracies of ResNet-18 (trained for 200 epochs with $|M_p|=200$ and $\mathcal{X}_p=\text{CIFAR10}$) under different $\epsilon$ are similar, indicating that data augmentation has little impact on the classification utility of InstaHide. Meanwhile, the increase of $\epsilon$ causes little impact on the proposed attack, since we filter the homogeneous encryptions to preserve those with limited $\epsilon$, from which we can restore the target images with high distinguishability and accurate color profiles.

\begin{figure*}[t]
\centering
\includegraphics[width=.7\textwidth]{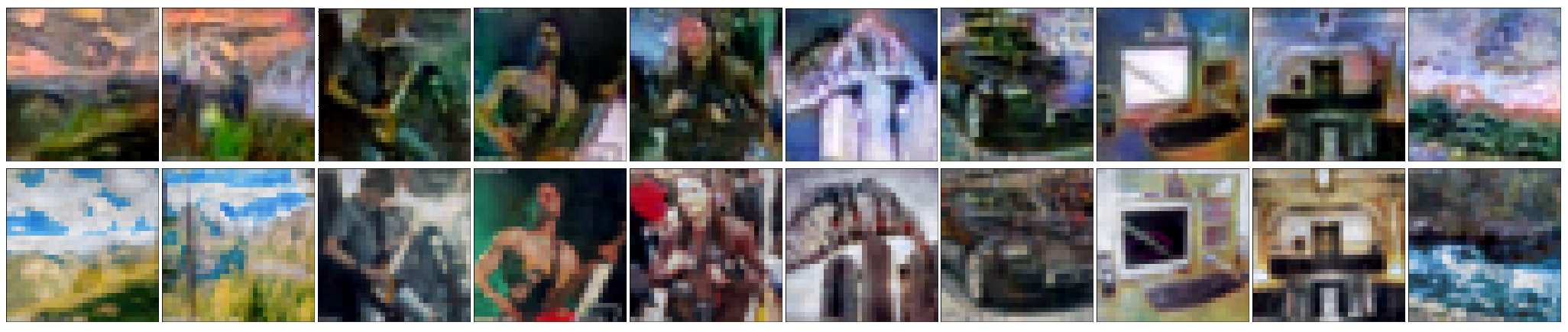}
\caption{Comparison between our attack (first row) and \cite{carlini2020attack} (second row) on InstaHide Challenge.}
\label{fig-google}
\end{figure*}

\vspace{1mm}
\noindent \textbf{Restored Images from InstaHide Challenge.}
We also employ our scheme to restore the private images from the InstaHide Challenge dataset~\cite{instachallenge}.
Since the private images in the InstaHide Challenge are not preprocessed by data augmentation, we thus train a modified \scheme\ with the image relaxing component replaced by a $3\times 3$ convolutional layer for preserving more details. The training dataset is CIFAR100, and $|M_t|$ is set to 50 since one private image is mixed into 50 encryptions in this dataset. We compare some images restored by our scheme and~\cite{carlini2020attack} in Fig.~\ref{fig-google}.
We observe that both our scheme and~\cite{carlini2020attack} can precisely restore the salient structures of the original images. However, the color profiles in some images restored by our scheme, such as the first and second images, are greatly different from the colors restored by~\cite{carlini2020attack}. Since the private images are not released, we can not compare the similarities between the restored images and the ground truth images. Nevertheless, the images restored by our scheme are visually recognizable, validating the effectiveness of \scheme.

\begin{figure*}[t]
\centering
\begin{subfigure}{.4\textwidth}
  \centering
  \includegraphics[width=0.95\textwidth]{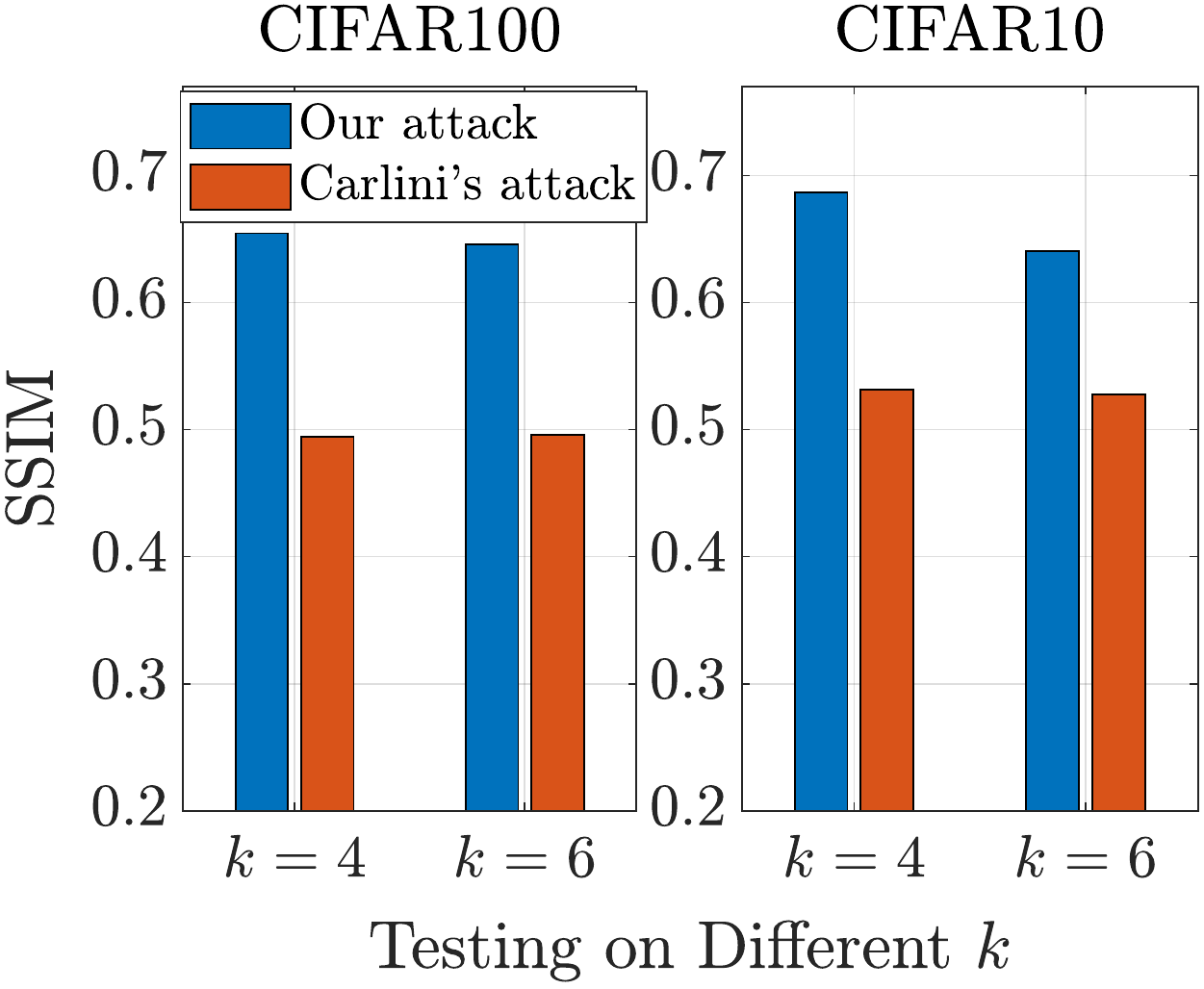}
  \caption{Performance}
  \label{fig-merge-k}
\end{subfigure}
~
\begin{subfigure}{.5\textwidth}
  \centering
  \includegraphics[width=0.95\textwidth]{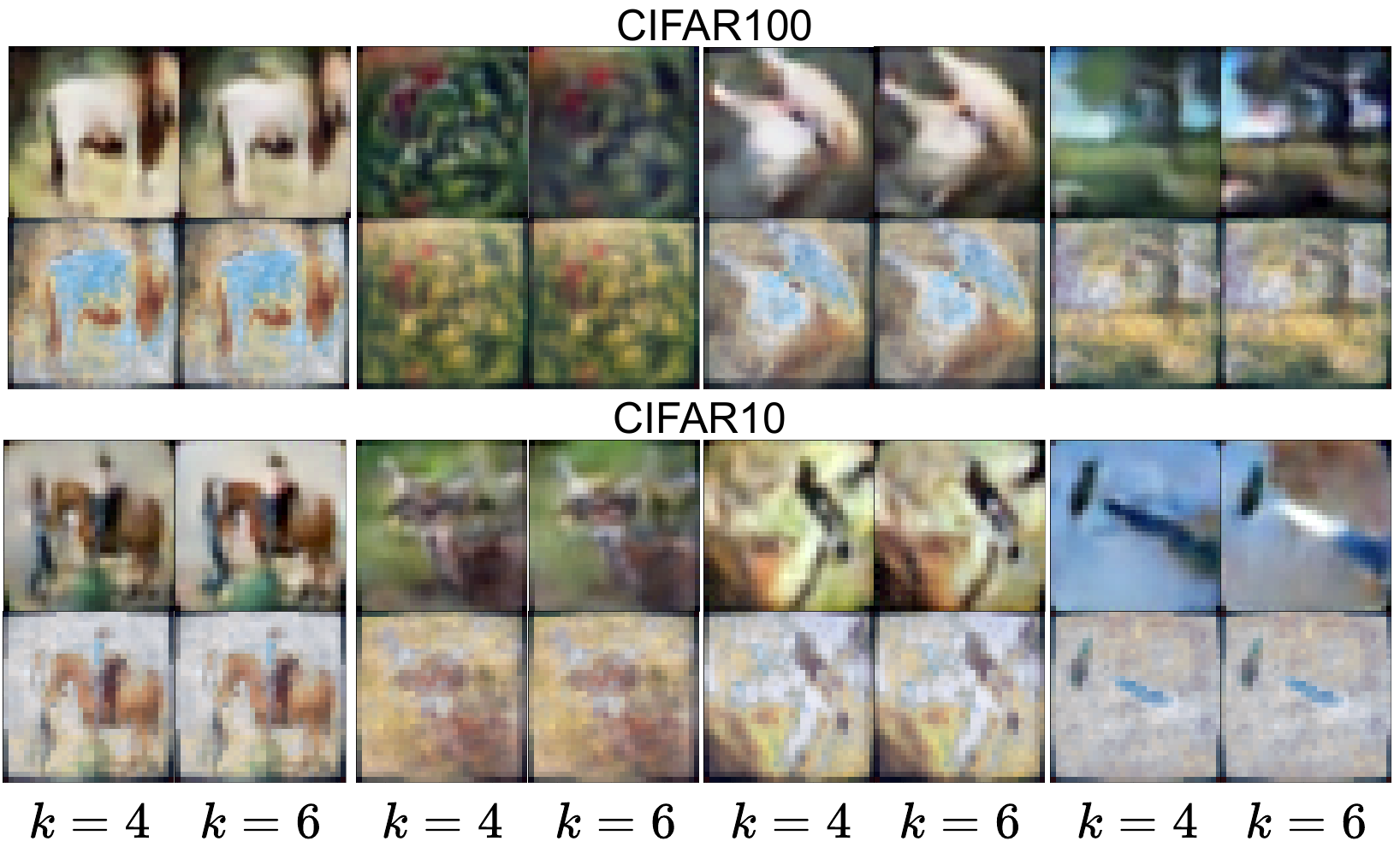}
  \caption{Examples \textit{w.r.t.} $k$}
  \label{fig-different-k}
\end{subfigure}
\caption{(a) the performance comparison between \scheme\ and CA-CN \textit{w.r.t.} different $k$; (b) the examples of images restored from InstaHide parameterized with $k=4$ and $k=6$. The first row shows images restored by our attack; the second row shows images restored by CA-CN.}
\end{figure*}

\vspace{1mm}
\noindent \textbf{Attack Performance under Different $k$.}
In~\cite{huang2020instahide}, the authors suggest two settings for the utility-security trade-off of InstaHide, i.e., $k=4$ and $k=6$. Compared to $k=4$, the InstaHide parameterized with $k=6$ mixes two more public images in one encryption, which introduces more noises and further corrupts the visual features of the private images.
In this section, we first fix $\mathcal{X}_t=\text{CIFAR100}$, $|M_p|=|M_t|=50$ and $\epsilon_p=\epsilon_t=0.1$, then train and test \scheme\ based on encryptions generated from $k=4$ and $k=6$, respectively. The results are shown in Fig.~\ref{fig-merge-k} and~\ref{fig-different-k}.

From Fig.~\ref{fig-merge-k}, we observe that the increase of $k$ slightly degrades the performance of \scheme\ while causing little impact on CA-CN. The reason is that the noise distribution in $k=6$ is more complicated than that of $k=4$, therefore being more difficult for the \scheme\ to learn. While \cite{carlini2020attack} uses gradient optimization to factor out all possible noises, which receives little effect from the changing of noise distributions. However, the performance of the proposed attack is still far better than that of~\cite{carlini2020attack}.
Fig.~\ref{fig-different-k} shows some examples of restored images from CIFAR10 and CIFAR100. We see that the increase of $k$ causes little visual impact on the restored images from both \scheme\ and CA-CN,
which indicates that mixing two more public images in one encryption could not greatly change the distribution of the mixed public components. In other words, \emph{the InstaHide parameterized with $k=6$ could not be more secure than the InstaHide parameterized with $k=4$}.

\begin{figure*}[htbp]
\centering
\includegraphics[width=.8\textwidth]{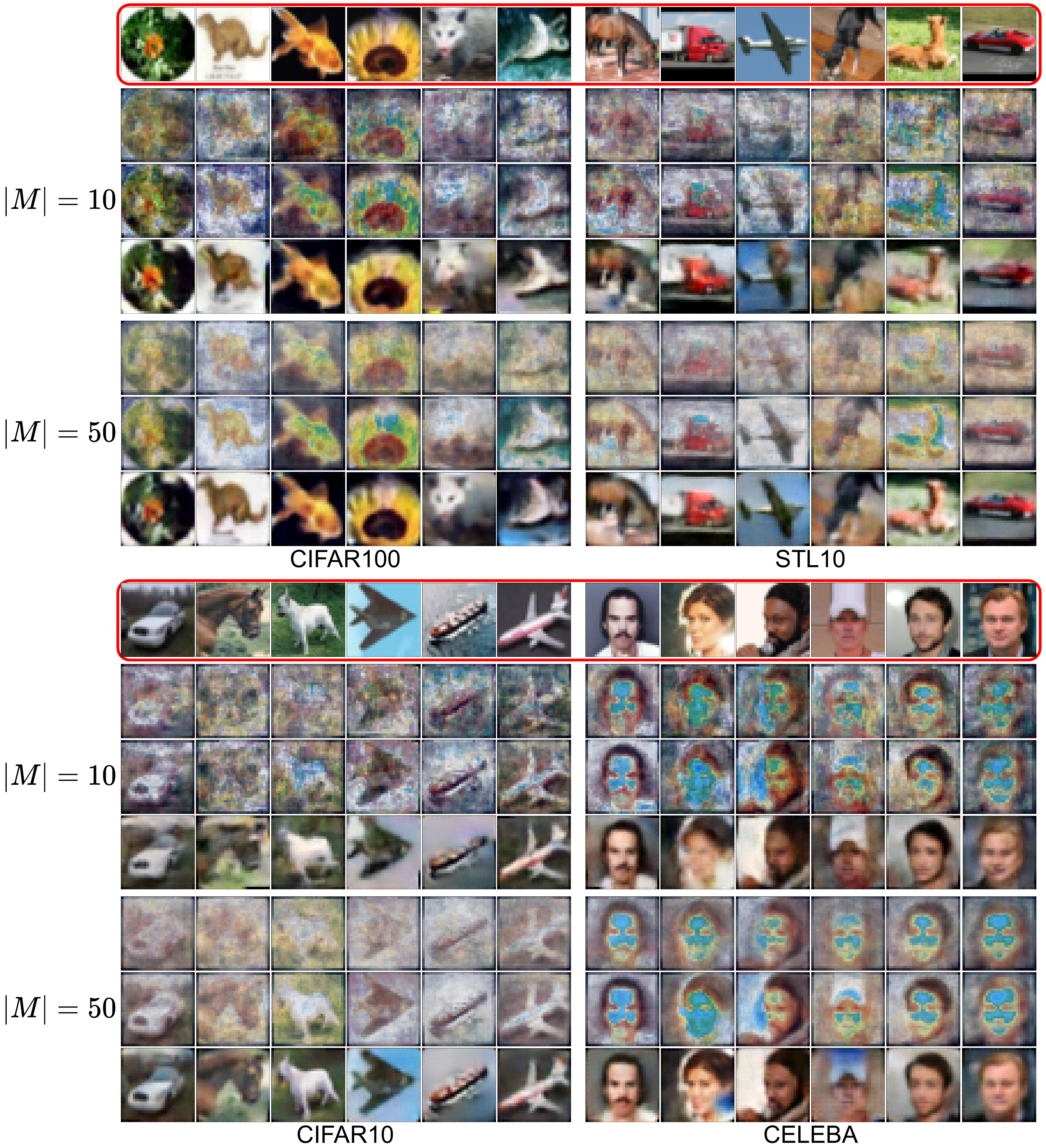}
\caption{Restored images from encryptions with different $|M_p|$ and $\epsilon_p=0.1$.
In each block, The first row shows images restored by CA; the second row shows images restored by CA-CN; and the third row shows images restored by \scheme. The images in the red box are the ground truth images.
}
\label{fig-differentM-fullexp}
\end{figure*}

\begin{figure*}[htbp]
\centering
\includegraphics[width=.8\textwidth]{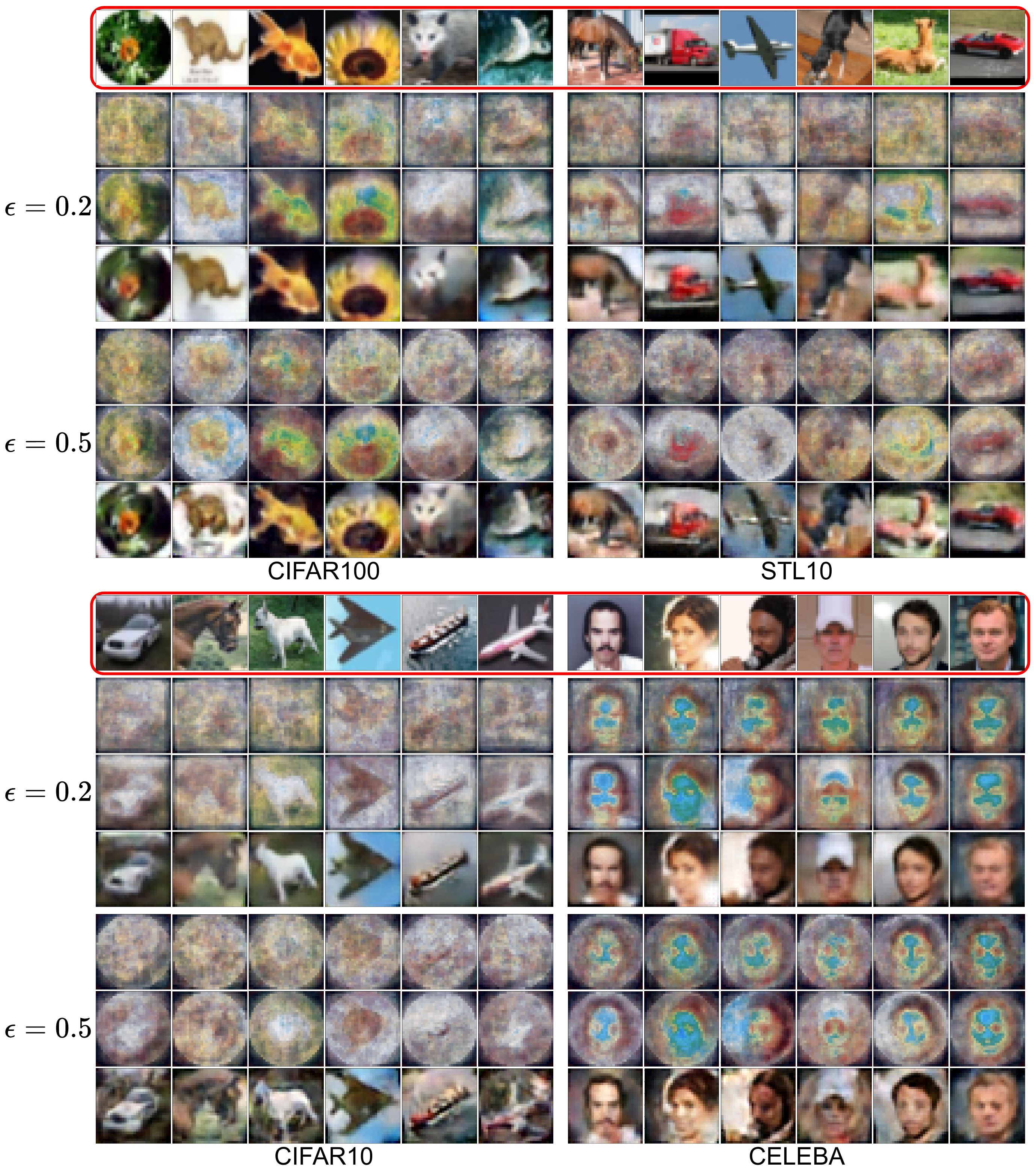}
\caption{Restored images from encryptions with different $\epsilon_p$ and $|M_p|=50$.
In each block, The first row shows images restored by CA; the second row shows images restored by CA-CN; and the third row shows images restored by \scheme. The images in the red box are the ground truth images.
}
\label{fig-differentexp-fullexp}
\end{figure*}

\vspace{1mm}
\noindent \textbf{More Results of the Experiment \emph{Comparison with Carlini et al.’s Attack}.}
We show more restored images in Fig.~\ref{fig-differentM-fullexp} and \ref{fig-differentexp-fullexp}. The corresponding experimental settings are given in the figure captions. 

\end{document}